\documentclass[11pt,a4paper]{article}

\usepackage[utf8]{inputenc}
\usepackage{tikz}
\usepackage{caption}
\usepackage{graphicx}
\usepackage{subcaption}
\usepackage[normalem]{ulem}
\usepackage{comment}
\usepackage{longtable}
\usepackage{pdfpages}
\usepackage{authblk}
\usepackage{amsthm}
\usepackage{thmtools}       
\usepackage[linesnumbered,ruled]{algorithm2e}
\usepackage{array}

\usepackage{booktabs}
\newcolumntype{L}[1]{>{\raggedright\arraybackslash}p{#1}}

\usepackage[
    backend=bibtex,
    style=numeric-comp,
    sorting=none,
    giveninits = true,
    doi = false,
    url = false,
    isbn = false,
    date = year,
    maxnames = 8
]{biblatex}
\bibliography{references}
\renewbibmacro{in:}{}
\newbibmacro{string+url}[1]{%
  \iffieldundef{url}{#1}{\href{\thefield{url}}{#1}}}
\DeclareFieldFormat{title}{\usebibmacro{string+url}{\mkbibemph{#1}}}
\DeclareFieldFormat[article]{title}{\usebibmacro{string+url}{\mkbibquote{#1}}}
\DeclareFieldFormat[misc]{title}{\usebibmacro{string+url}{\mkbibquote{#1}}}
\DeclareFieldFormat[inproceedings]{title}{\usebibmacro{string+url}{\mkbibquote{#1}}}
\DeclareFieldFormat[incollection]{title}{\usebibmacro{string+url}{\mkbibquote{#1}}}
\DeclareFieldFormat[book]{title}{\usebibmacro{string+url}{\mkbibquote{#1}}}

\usepackage{amsmath}
\usepackage{quantum}

\usepackage{listings}
\usepackage{amssymb}
\usepackage[ampersand]{easylist}
\usepackage{enumitem}
\usepackage{hyperref}
\hypersetup{
  colorlinks = true, 
  linkcolor={blue!50!black}, 
  citecolor={blue!50!black},
  urlcolor=blue, 
  linktoc=page,
}
\usepackage[capitalise, nameinlink]{cleveref}
\usepackage[
  style=long,
  nolist, 
  nonumberlist,
  nopostdot,
  acronym,
  toc=true, 
]{glossaries}
\usepackage{bm}
\usepackage{booktabs}
\usepackage{multirow}

\usepackage[british]{babel}
\usepackage{csquotes}

\title{Enhancing density functional theory using the variational quantum eigensolver}

\author[1]{Evan Sheridan}
\author[1]{Lana Mineh}
\author[1]{Raul A. Santos}
\author[1]{Toby Cubitt}
\affil[1]{Phasecraft Ltd.}

\date{\today}

\begin{document}

\maketitle

\begin{abstract}
  Quantum computers open up new avenues for modelling the physical properties of
  materials and molecules. Density Functional Theory (DFT) is the gold standard
  classical algorithm for predicting these properties, but relies on
  approximations of the unknown universal functional, limiting its general
  applicability for many fundamental and technologically relevant systems. In
  this work we develop a hybrid quantum/classical algorithm called \emph{quantum
    enhanced DFT} (QEDFT) that systematically
  constructs quantum approximations of the universal functional using data
  obtained from a quantum computer.

  We benchmark the QEDFT algorithm on
  the Fermi-Hubbard model, both numerically and on data from experiments on real quantum hardware. We find that QEDFT surpasses the quality of
  groundstate results obtained from Hartree-Fock DFT, as well as
  from direct application of conventional quantum algorithms such as VQE.
  Furthermore, we demonstrate that QEDFT works even when only noisy, low-depth quantum computation is available, by benchmarking the algorithm on data obtained from Google's quantum computer.

  We further show how QEDFT also captures
  quintessential properties of strongly correlated Mott physics for large
  Fermi-Hubbard systems using functionals generated on much smaller system
  sizes. Our results indicate that QEDFT can be applied to realistic
  materials and molecular systems, and has the potential to outperform the direct application of either DFT or VQE alone, without the requirement of large scale or fully fault-tolerant
  quantum computers.
\end{abstract}

\tableofcontents

\section{Introduction}
Materials systems constitute large, interacting and inhomogeneous quantum many-body problems.
As such, understanding and describing their properties is a notoriously difficult problem.
Exact simulation of materials on classical computers is restricted to small system sizes.
Quantum simulation on quantum computers promises precise computation of their dynamics, but remains out of reach of current hardware~\cite{clinton2022towards}.
Instead, a number of well reasoned approximations across different time and length scales are used to make consistently reliable predictions.

\subsection{Density Functional Theory}
At the atomic scale, Density Functional Theory (DFT)~\cite{DFTI, DFTII} is the predominant computational method used for computing groundstate electronic and structural properties for a broad range of atomic, molecular and materials systems.
DFT computations have their basis in the Hohenberg-Kohn~\cite{DFTI} and Kohn-Sham~\cite{DFTII} theorems, which show that for any electronic structure Hamiltonian, there always exists a non-interacting Hamiltonian from which a groundstate density can be constructed that is identical to that of the full many-body system.
Of course, the local -- also known as Kohn-Sham (KS) -- potential in this non-interacting Hamiltonian is itself determined by the many-body groundstate density. So finding the correct KS potential for a given system is at least as hard as the original many-body problem.
In DFT, a way of getting around this is to approximate the energy, as a functional of the density. This energy functional $E[n]$, where $n$ is the groundstate density, decomposes into four parts:
\begin{equation}
  E[n] = T[n] + U_{H}[n]  + E_{\text{ext}}[n] + E_{\mathrm{xc}}[n].
\end{equation}
The kinetic energy term $T[n]$, the Coulomb (or ``Hartree'') term $U_{H}[n]$, the external potential $E_{\text{ext}}[n]$,  and the ``exchange-correlation'' (XC) term $E_{\mathrm{xc}}[n]$.
This latter term is essentially defined as everything not accounted for by the other terms:
\begin{equation}
  E_{\mathrm{xc}}[n] = E[n] - T[n] - U_{H}[n] - E_{\text{ext}}[n].
\end{equation}

The kinetic, Hartree, and external terms can be computed classically within KS DFT~\cite{DFTII}.
Meanwhile, $E_{\mathrm{xc}}[n]$ only depends on the electron-electron interactions, so in principle is a universal functional that applies to every many-body electronic structure problem.
However, since knowledge of this universal functional would allow all many-body problems to be solved, determining the exact form of this universal functional is believed to be intractable~\cite{DFT_QI, universal_review}.

Instead, in DFT, various heuristic approximations of the XC functional are used.
Simple and compact approximations are often the most useful, interpretable and computationally efficient~\cite{dft_func}.
In its most common formulation, the algorithmic complexity of DFT scales as $\mathcal{O}(N^3)$, with $N$ being the number of particles, whereas the full quantum many-body simulation scales exponentially in the particle number on classical computers.

However, the performance of DFT is intrinsically limited by the accuracy of the XC functional approximation.
The most basic and best-known of these is the local density approximation (LDA), which is determined by exact computation of the homogeneous, non-interacting electron gas~\cite{LDA_QMC}.
Other more sophisticated parameterisations exist and are used in modern DFT computations~\cite{dft_func} with different levels of classical algorithmic complexity.
Over more than 50 years, hundreds of approximate XC energy functionals have been proposed~\cite{Libxc}, for both continuous basis sets as well as model Hamiltonians~\cite{LDFT_BALDA}, and used in virtually every facet of atomistic materials modelling.
This has resulted in the foundational papers of DFT becoming some of the most cited works in the scientific literature~\cite{DFT_citations, DFT_RMP} and the birth of the field of ab-initio materials design~\cite{Marzari2021}.

Despite these many successes, a number of challenges still remain: from choosing the most appropriate functional for each system, to the general failure of XC functionals when applied to strongly correlated systems.
There are many functionals which have impressive records across broad material classes when predicting certain properties, while completely failing for others.
For example, many functionals tend to struggle with describing bond disassociation: when electrons localise on ions as the inter-atomic distance is increased. And this is closely related to the misprediction of lattice constants in solids even when the electronic structure is qualitatively correct~\cite{lattice_under}.
Moreover, for strongly correlated materials systems, the current generation of XC potentials often make qualitatively bad predictions for electronic and structural groundstate properties, let alone attaining quantitatively accurate results.

In recent years, quantum many body approaches such as the Dynamical Mean Field Theory (DMFT)~\cite{DMFTI_RMP} and quasiparticle self-consistent GW~\cite{qsgw} approaches have alleviated these issues for some strongly correlated systems.
In these approaches, the XC functional is not modified or updated; instead, exact quantum simulations for restricted regions of the Hilbert space are performed multiple times.
However, this requires extensive classical computational power for each new material.

\subsection{Quantum computation} \label{sec:vqe}
Recently, quantum computing hardware has become sufficiently advanced to allow non-trivial demonstrations of quantum computation of groundstate properties.
Many different quantum algorithms have been proposed for finding groundstates.
However, the most widely studied on near-term quantum hardware is the variational quantum eigensolver (VQE)~\cite{vqeI, vqeII}.
At its heart, like many classical variational algorithms, VQE is based on the Rayleigh-Ritz variational characterisation of the eigenvalues of the Hamiltonian $H$:
\begin{equation}
  \lambda_0(H) = \min_{\ket{\psi}} \braket{\psi|H|\psi}.
\end{equation}
However, in contrast with classical variational algorithms, in VQE the state $\ket{\psi}$ is generated --- and its energy $E(\ket{\psi}) = \braket{\psi|H|\psi}$ with respect to the Hamiltonian $H$ measured --- on a quantum computer.
This requires that a suitable, efficiently preparable class of quantum states $\ket{\psi(\vec{p})}$ be chosen, with parameters $\vec{p}$ over which to optimise.
Because quantum computers are able to efficiently construct states that cannot be described or computed efficiently on classical computers, VQE has the potential to outperform classical variational methods.

The accuracy of VQE results depends on whether this variational class contains a good approximation to the groundstate, and many different variational classes have been proposed and studied~\cite{Cerezo2021, Bharti2022}.
The performance of VQE also depends on the success of the classical optimisation routine used to optimise the measured energy $E(\ket{\psi(\vec{p})})$ over the parameters $\vec{p}$.
Theoretically, good choices of the variational class and classical outer optimiser show promise in solving certain quantum many-body problems~\cite{Tilly2022}.

However, the current regime of noisy intermediate-scale quantum (NISQ) hardware is notoriously subject to significant rates of noise and error.
The success or otherwise of VQE therefore also depends, crucially, on hardware noise.
Although non-trivial VQE computations have now been demonstrated on real quantum devices~\cite{stanisic2022observing}, none have yet gone beyond system sizes that can be simulated on classical computers.
Although the results reproduce the correct physical properties in these cases, the VQE \emph{state} itself, as generated on the quantum computer, is not necessarily close to the true groundstate.

\subsection{Quantum enhanced DFT}
In this work, we introduce and benchmark a new hybrid quantum/classical approach to the many-body electronic structure problem: \emph{quantum-enhanced DFT} (QEDFT), which combines the complementary strengths of DFT and quantum computation.
Rather than using the quantum computer to find the groundstate itself, we instead use the quantum computer to approximate the XC functional, which is then fed into a classical DFT iteration.
Thus the quantum computer is not tasked with finding the groundstate itself, but rather with steering the DFT iteration towards an accurate solution.
In this work, we focus on using VQE for the quantum part of the QEDFT computation.
But we note that our approach is general, and any other quantum groundstate approximation algorithm could be used in QEDFT.

We find that combining DFT and VQE in this way outperforms both Hartree-Fock DFT alone, and quantum VQE alone.
Moreover, this remains the case even when the VQE computation itself is very noisy, and its output is far from the true XC functional.
Despite the VQE output being quantitatively inaccurate, it captures key qualitative features of the exact XC functional that classical approximations struggle to reproduce.
And this suffices to steer the DFT iteration to more accurate solutions than it would otherwise converge to.

To carry this out, the evaluation of the XC functional must be transformed into a groundstate computation suitable for VQE.
At a high level, there are two possible approaches: we can compute the XC functional via VQE during the DFT loop, at the density in the current DFT iteration; we call this the \emph{online} approach.
Or we can use VQE to approximate the entire universal XC functional in advance, and then use this VQE-computed XC functional in the DFT iteration; we call this the \emph{offline} approach.
There are pros and cons to each approach: the online approach only evaluates the XC functional at densities actually required during the DFT iteration; whereas the offline approach may compute it at many densities that are not actually needed.
However, the online approach requires calling out to the quantum computer during the DFT computation, thereby limiting the speed of the DFT computation; whereas the offline approach decouples the quantum and classical computation.
Moreover, since we are approximating the universal functional, in principle the VQE functional approximation only needs to be computed once, and thereafter the same functional can be used in multiple QEDFT computations.

For large systems, the XC functional is essentially a continuous object, whereas a quantum computer can only compute (or approximate) its value at a discrete set of points, with intermediate points approximated by interpolation.
In principle, assuming the VQE algorithm were able to compute the exact value of the functional at any given point, the finer the discretisation the more accurate the functional approximation.
However, the fact that a discretised functional is being computed has a significant benefit.
The number of points in the discretisation corresponds directly to the size of the quantum system used in the quantum computation to compute the value of the functional.
Therefore, even if the quantum hardware is too small to fit the full system being simulated, we can still use VQE to approximate the XC functional on the quantum computer at the finest discretisation that will fit on the device, and then use this in a DFT calculation of the groundstate density of the much larger system of interest.

Our empirical study shows that QEDFT has a number of advantages over both Hartree-Fock DFT alone, and VQE (or other purely-quantum) approaches alone:
\begin{enumerate}
\item QEDFT usually achieves better accuracy than either Hartree-Fock DFT alone, or quantum VQE alone.
\item QEDFT does not necessarily rely on quantitatively accurate quantum computations (eg high fidelity with the true groundstate), so it may provide a path to quantum advantage even when the quantum hardware is noisy.
\item QEDFT can be applied to large many-body systems on small quantum hardware, so it may still be able to provide a quantum advantage for real materials even while the available hardware is too small to fit a full quantum simulation of the material.
\end{enumerate}

All of the numerics and hardware experiments across 1D and 2D Fermi-Hubbard
systems serve to support these claims. We highlight a subset of our
results that illustrate the above claims. \Cref{fig:fig6} shows that QEDFT achieves better accuracy than DFT Hartree Fock on the 1D Fermi-Hubbard chain.
\Cref{fig:fig7} support the claim that QEDFT succeeds even when the output of the quantum part of the computation is noisy and quantitatively inaccurate. \Cref{fig:fig6} and \Cref{fig:fig9} show results obtained for large 1D and 2D Fermi-Hubbard lattices using small quantum computations.

\subsection{Previous work}\label{section:quantum_review}\label{section:classical_methods}
The first studies of DFT and quantum computing analysed its computational complexity~\cite{DFT_QI}. However, with the advent of readily available near-term devices, there have been a number of early studies of the development of hybrid quantum-DFT algorithms being applied to small molecules~\cite{rossmannek2021quantum} and mid-gap defect states in solid state materials~\cite{ma_quantum_2020, galli22, galli22_II}.
In these works, DFT is bootstrapped by VQE~\cite{vqeI, vqeII} and quantum subspace expansions~\cite{QSE} by active space embedding methods, where the full KS space of electrons is partitioned into classical and quantum sections.
There is also a formulation in the spirit of DFT in the Zumbach Maschke basis~\cite{hatcher_method_2019} that outlines a theoretical proposal to compute an approximation of the XC potential based on density matrix measurements that iteratively correct the single-particle density towards the true solution.
For the fault tolerant framework, an approach has been proposed that could obtain accurate approximations to the XC functional that uses machine learning~\cite{baker_density_2020}, but requires the use of quantum algorithms not suited for near-term devices.

Finally, two recent works~\cite{Desh_qdft, senjean2022toward} apply the formalism of DFT to the 1D Hubbard model using quantum algorithms.
In~\cite{Desh_qdft} the authors use VQE within the Levy-Lieb formalism of DFT to compute the groundstate properties of the Hubbard dimer, supplementing their technique with a quantum kernel that can learn the density functionals of observables.
On the other hand, in~\cite{senjean2022toward}, a mapping between the KS Hamiltonian and an auxiliary Hamiltonian is implemented on a quantum computer. The authors posit that solving the auxiliary system instead can result in computational speed up.
This approach is then applied to 1D Hubbard chain of size $L=8$ and a hydrogen molecular chain.

We also note that functional design for correlated lattice models has also been explored classically in the context of various lattice inhomogeneities.
It was first used in semiconductor models~\cite{LDFT_original}, but it has since been more routinely applied to one dimensional Hubbard chains~\cite{LDFT_BALDA, xianlong2006bethe, CAPELLE201391, carrascal2015hubbard, LDFTII}, where exact parameterisations of the XC functional are available based on the Bethe Ansatz.
In 2D, this approach has highlighted groundstate properties of large inhomogeneous sheets of graphene in a tight binding model with Hubbard type interactions, where the exact functional was parameterised using exact diagonalisation data~\cite{LDFT_graphene}.
Additionally, in recent years machine learning methods have been trained against exact results, showing utility for parameterising the XC functional for molecules~\cite{burkeML, BurkeMLII, DM21} and also for the Hubbard model~\cite{sanvitoML}.
It is now clear that improving methods for XC functional design is a key factor in unlocking the full range of capabilities afforded by KS DFT.

\subsection{Outline}
In the rest of this paper, we describe the QEDFT algorithm in more technical detail, and apply it to a number of many-body lattice models.
Lattice models are good systems to benchmark the approach on, as they allow direct comparison to other quantum algorithms which have largely been tested in the lattice setting.
Furthermore, because quantum computers consist of qubits, fundamentally any quantum simulation must be reduced in some way to a discretised model.
Thus a lattice formulation of QEDFT is required in any case, whether explicitly or implicitly.

In \cref{sec:lattice}, we review the formulation of DFT on lattices, which differs slightly from the more familiar continuum formulation of DFT.
In \cref{sec:QEDFT}, we describe how the XC functional is transformed into a groundstate problem amenable to quantum computation.
We also explain how a quantum treatment on small system sizes can be used to interpolate approximations of the XC functional, which can then be used in DFT for much larger system sizes.
As a simple illustrative example, we first show in \cref{sec:examples} how our QEDFT approach looks for the Hubbard dimer and then extend this to more complex Fermi-Hubbard models.
In \cref{sec:numerics}, we compare the performance of the QEDFT algorithm numerically with both (simulated) VQE, and with DFT under a number of choices of classical functional approximations.
We also explore the qualitative features of the QEDFT functional and their contribution to the algorithm's performance.
In \cref{sec:real_quantum_hardware}, we run the QEDFT algorithm on the 1D Fermi-Hubbard model using VQE data obtained on Google's quantum device.

\section{Lattice DFT}\label{sec:lattice}


We consider the general Hamiltonian for the following many-body fermionic system
defined on a lattice $\mathcal{L}$,
\begin{equation}
  H = \hat{T} + \hat{H}_{\text{int}} + \hat{H}_{\text{ext}},
  \label{eq:DFT_gen}
\end{equation}
where $\hat{T}$,  $\hat{H}_{\text{int}}$ and $\hat{H}_{\text{ext}}$ are the non-interacting, interacting and external field
contributions respectively. DFT enables the efficient solution of this problem
\cite{LDFT_original}, under the assumptions of non-interacting
$\nu$-representability of an auxiliary system to \cref{eq:DFT_gen} and the linear coupling of
$\hat{H}_{\text{ext}}$ to the local density, using density functionals rather
than wavefunctions. The validity of the DFT approach has applicability for any
Hamiltonian which linearly couples to an external field, which for example could
be the electric field, but more generally takes the form
\begin{equation}
  H_{\text{ext}} = \sum_{i\in \mathcal{L}} v^{ext}_i \hat{n}_i,
\end{equation}
where $v^{ext}_i$ is the external potential. The first Hohenberg-Kohn (HK)
theorem establishes a one-to-one mapping~\cite{DFTI} between the groundstate
electron density, $n_{i} = \langle \psi_{GS}| \hat{n}_i| \psi_{GS}\rangle$, and
the many body groundstate wavefunction, $|\psi_{GS}[n]\rangle$, which is a
functional of the density. The second HK theorem states that the groundstate
electron density minimises the energy functional~\cite{DFTI}. It follows that the
expectation value of any observable in the groundstate is uniquely defined as a
functional of the groundstate density. For example, the groundstate energy
functional is given by
\begin{equation}
  E_{GS} = F[n] + \sum_{i\in \mathcal{L}} v^{ext}_i n_i,
  \label{eq:uni_energy}
\end{equation}
where $F[n] = \langle \psi_{GS}[n] | \hat{T} + \hat{H}_{\text{int}} |
\psi_{GS}[n] \rangle$ is known as the universal functional, as it doesn't depend
explicitly on the external potential. The
exact analytical form of this functional is not known for many fermionic systems
of interest and requires the solution of many body
problems that do not scale efficiently in the system size on classical
computers. Routine approximations of $F[n]$ are made to allow for practical
implementations of DFT. The error in the estimation of groundstate observables depends on the quality of those approximations.

A necessary reduction is then made from the original problem to an auxiliary
non-interacting (KS) Hamiltonian~\cite{DFTII} that is guaranteed to
reproduce the groundstate properties of the original system only if the exact
universal functional is used. This step is valid only if the original system is
non-interacting $\nu$-representable, i.e the groundstate density of the
original interacting system can be obtained from a non-interacting one. Within
this scheme, known as the KS approach, \cref{eq:uni_energy} is written
as,
\begin{equation}
  E_{GS}[n] = T_{NI}[n] + U_{H}[n] + E_{XC}[n] + \sum_{i\in \mathcal{L}} v^{ext}_i n_i,
  \label{eq:uni_energy_II}
\end{equation}
where $T_{NI}[n]$ is the kinetic energy of the auxiliary non-interacting system,
$U_H[n]$ is the Hartree energy, and $E_{XC}[n]$ is the exchange-correlation
energy functional. By minimising \cref{eq:uni_energy_II}, the Euler-Lagrange
equations result in,
\begin{equation}
  \frac{\partial E_{GS}}{\partial n_i} = \frac{\partial T_{NI}}{\partial n_i} + \frac{\partial U_{H}}{\partial n_i} + \frac{\partial E_{XC}}{\partial n_i} + v^{ext}_i = 0,
\end{equation}
which has a solution that is the groundstate density. This equation is
equivalent to a non-interacting system with an effective potential defined as,
\begin{align}
  V^{\text{KS}}_i &= v^{ext}_i + \frac{\partial U_H}{ \partial n_i} + \frac{\partial E_{XC}}{\partial n_i} \nonumber \\
  &= v^{ext}_i + V_{H}[n_i] + V_{XC}[n_i],
  \label{eq:KS_potential}
\end{align}
where $V_H[n_i]$ is the Hartree potential and $V_{XC}[n_i]$ is the exchange
correlation potential, which is the object that is typically approximated. The
effective non-interacting single-particle system is known as the KS
Hamiltonian,
\begin{equation}
  \hat{H}_{KS} = \hat{T} + \sum_{i\in \mathcal{L}}  V^{\text{KS}}_i \hat{n}_i.
  \label{eq:KS_eqns}
\end{equation}

This system of equations is then solved self-consistently where at each
iteration the groundstate density at site index $i$ is constructed from the
eigenvectors of the KS system, by summing over the occupied single particle states of the
KS eigenvectors (accounting for spin),
\begin{equation}
  n_{GS,i} = 2 \sum_{j}^{\text{occ}/2} |\phi_{j}(i)|^2,
  \label{eq:KS_density}
\end{equation}
where $\hat{H}_{KS} \phi_j = \epsilon_j \phi_j$ is the full eigensolution of the KS
system. Here we treat the example spin unpolarised lattice DFT, but we note that spin polarised
lattice DFT can be treated by separating the Kohn-Sham states into their up and down spin channels.
The groundstate energy of the original many-body system in terms of the KS eigenvalues and density is given by,
\begin{equation}
  E_{GS}[n_{GS}] = 2\sum_{i \leq \text{occ}/2} \epsilon_i - \sum_{i\in \mathcal{L}} V_{XC}[n_{GS,i}]n_{GS,i} - E_{H}[n_{GS}] + E_{XC}[n_{GS}].
  \label{eq:gs_energy}
\end{equation}

The KS solution is converged self-consistently by iteratively updating
the input density by computing \cref{eq:KS_density} and mixing it with the previous
density estimate,

\begin{equation}
  n^{(j+1)} = \alpha n^{(j)} + (1 - \alpha)n^{(j+1)}
  \label{eq:mixing}
\end{equation}
where $j$ is the DFT iteration index and $\alpha$ is a linear mixing parameter
that is used for numerical stability. When the groundstate observables no
longer change up to a tolerance $\delta$ then the groundstate energy and
density are equivalent to those of the interacting problem, if the
true XC energy functional is used.


\begin{table}[!htb]
  \centering
      {\LinesNumberedHidden
        \begin{algorithm}[H]
          \SetKwInOut{Input}{Input}
          \SetKwInOut{Output}{Output}
          \SetAlgorithmName{Algorithm}{}
          CChoose the inhomogeneous representation (i.e the external potential) of the fermionic system to be used \cref{eq:DFT_gen} 
          \begin{enumerate}
          \item Choose initial density;
          \item Construct the KS potential \cref{eq:KS_potential} where $V_{XC}$ is approximated;
          \item Solve for $(\phi_j,\epsilon_j)$ in the KS system of equations $\hat{H}_{KS}\phi_j=\epsilon_j\phi_j$ with $\hat{H}_{KS}$ \cref{eq:KS_eqns};
          \item Construct density estimate from the KS eigenvectors \cref{eq:KS_density};
          \item Compute energy of the system \cref{eq:gs_energy};
          \item Check for system convergence;
          \item If converged, exit loop and compute groundstate properties. If not, go back to step (1) and use new estimate of density as the input density \cref{eq:mixing};
          \end{enumerate}
          \caption{The lattice DFT algorithm}
      \end{algorithm}}
      \caption{The algorithmic procedure for executing a DFT loop for practical applications. }
      \label{algo:DFT_loop}
\end{table}

To summarise, the groundstate density minimises the total energy functional
which enforces the functional derivative to be zero and results in a variational
equation for auxiliary wavefunctions that can construct the groundstate
density. This equation takes the form of a single-particle equation which can be
treated as a standard eigenvalue problem. All of many-body physics manifests in
the exchange correlation potential contribution of an otherwise fully determined
effective single particle potential. These equations are solved self
consistently, as the XC potential depends on the density, and the density
depends on the unknown auxiliary wavefunctions. At self consistency, these wavefunctions
no longer change and are used to construct groundstate observables, most
notably the density and energy. The algorithmic procedure is presented in \cref{algo:DFT_loop}.

\section{The QEDFT algorithm}\label{sec:QEDFT}

In this section we present the quantum-enhanced DFT algorithm for fermionic
systems within the lattice DFT formalism presented in \cref{sec:lattice}. The main difference between
lattice DFT and normal (ab initio) DFT is that the former is discrete, while the latter
uses continuous basis sets.
The algorithm consist of two parts (a) the computation of the XC potential and energy
using a quantum machine and (b) application of the potential and energy within a
DFT loop, which essentially follows the steps outlined in
\cref{sec:lattice}. We discuss QEDFT in its offline mode, where the quantum
computer is called during the first stage only and the second stage is a purely
classical computation that queries that results from the first stage. Our
approach is motivated by the deficiencies of classical DFT approximations to
systematically incorporate the effects of exchange and correlations. On the
other hand, quantum computations natively include these effects, so building
functionals out of their computations has the potential to improve upon
classical approximations.

In \cref{algo:XC_potential} we describe the algorithmic procedure for
computation of the quantum-enhanced XC (QE-XC) potential and energy in this scheme. The initial step
of the problem is to choose a homogeneous system of size $L$. The homogeneous
Hamiltonian is related to the inhomogeneous Hamiltonian of \cref{eq:DFT_gen} by
setting $\hat{H}_{\text{ext}}=0$. Its groundstate energy for density
$n=\{n_1, \hdots, n_L\}$ is given by the universal functional $F[n=\{n_1,
  \hdots, n_L\}]$, where $\sum_i n_i = N_e$ corresponds to the total number of
electrons in the system.
The fermionic Hamiltonian of $L$ sites is then encoded into a
$M=\mathcal{O}(2L)$ qubits Hamiltonian via a fermion-to-qubit mapping,
\begin{equation}
  H^{\text{fermions}} = \sum_k h_k \rightarrow H^{\text{qubits}} = \sum_{\alpha} \lambda_{\alpha} P_{\alpha},
  \label{eq:JW_mapping}
\end{equation}
where $P_{\alpha} \in \{ I, X, Y, Z \}^{\otimes M}$ are Pauli operators and
 $\lambda_{\alpha}$ are real coefficients. This mapping is necessary due to how operations are executed on the
quantum device and its choice depends on the system under consideration.

A quantum circuit representation $|\Psi_{QC}[n]\rangle$ is then created that can be used by quantum
algorithms to predict the groundstate energy of the system for a fixed
number of fermions. The variational quantum eigensolver (\cref{sec:vqe}) can be used for this
step, where a quantum circuit is parameterised by real parameters that are
variationally optimised for. At each allowed fermionic filling, the quantum
algorithm is run to compute the groundstate energy,
\begin{equation}
  E(n)=\langle \Psi_{QC}[n]|H^{\rm qubits}|\Psi_{QC}[n]\rangle\,\, \text{ where } \sum_{i}n_i = N_e, \hfill \quad \forall N_e \in  \{ 0, \hdots, 2L\}.
  \label{eq:vqe_loop}
\end{equation}
At this stage, there is still the combinatorially many possibilities for the densities at each site of the lattice $\mathcal{L}$ as $n=\{n_1,\dots,n_L\}$. To further reduce the complexity we employ the local density approximation, where we just scan over states with total density $N_e/L$. This naturally fits in the state preparation of $|\Psi[n]\rangle$ as the total density is a conserved quantity, so the VQE minimization rearranges the local density profile while maintaining $N_e$.

From now on, within the offline QEDFT mode all computations are executed on a
classical machine that post-processes the function $E(n)$ with a spacing $1/2L$.
We introduce the quantum-enhanced local density approximation $\epsilon_{QELDA}$ for the energy \cref{eq:vqe_loop},

\begin{equation}
  \epsilon_{QELDA}\left(\frac{N_e}{L}\right) = \frac{E(n)}{L},\quad \text{s.t }\quad \sum_i n_i = N_e,
  \label{eq:LDA_def}
\end{equation}
over the domain $\frac{N_e}{L}\in [0,2]$. Therefore, as the system size is increased, the number of grid points that can
be be accessed by $\epsilon_{QELDA}(\frac{N_e}{L})$ increases and the resolution
of the local energy increases. The local XC energy can also be obtained via,
\begin{equation}
  \epsilon_{QELDA-XC}\left(\frac{N_e}{L}\right) =   \epsilon_{QELDA}\left(\frac{N_e}{L}\right) - \epsilon_{HF}\left(\frac{N_e}{L}\right),
  \label{eq:local_XC_energy}
\end{equation}
where $\epsilon_{HF}(\frac{N_e}{L})$ is the local Hartree Fock (HF) energy which can
be obtained efficiently on a classical machine. In this representation, the total
exchange correlation energy is approximated as a linear
combination of the local energy generated by the local density at each site, i.e.
\begin{equation}
  E_{QE-XC}(n) = \sum_{i\in\mathcal{L}} \epsilon_{QELDA-XC}(n_i).
  \label{eq:E_QEXC}
\end{equation}

The QE-XC potential is then obtained by taking the numerical derivative of
\cref{eq:E_QEXC},
\begin{equation}
  V_{QE-XC}(n_i) = \left.\frac{\partial E_{QE-XC}(n)}{\partial n_i} \right|_{n=n_{GS}}.
  \label{eq:QEXC_potential}
\end{equation}

As \cref{eq:QEXC_potential} and \cref{eq:local_XC_energy} are only known in the
domain $[0,2]$ with grid spacing $1/2L$  
it is necessary to generate a continuous representation of it so that it can be queried in a DFT loop. This can
be achieved by performing a spline interpolation for a fixed polynomial order.
As a result of this interpolation, the functional can be used for target systems
which are not the ones used to generate it. For example, a functional which is
generated on a 2D system of size $3\times3$ can be used for a
target system of size $30 \times 30$. The DFT loop then proceeds as described in
\cref{sec:lattice} for any Hamiltonian system given by \cref{eq:DFT_gen}, where the
form of the XC potential and energy used in the DFT query is given by the continuous
representations of \cref{eq:E_QEXC} and \cref{eq:local_XC_energy}.

\begin{table}[!htb]
  \centering
      {\LinesNumberedHidden
        \begin{algorithm}[H]
          \SetKwInOut{Input}{Input}
          \SetKwInOut{Output}{Output}
          \SetAlgorithmName{Algorithm}{}
           CChoose system of interest and construct its homogeneous fermionic Hamiltonian representation, i.e \cref{eq:DFT_gen} where $\hat{H}_{\text{ext}}=0$;
          \begin{enumerate}
          \item Encode the fermionic Hamiltonian into qubits via a fermion-to-qubit mapping \cref{eq:JW_mapping};
          \item Create a quantum circuit representation for computation of groundstate energy;
          \item Use a quantum computer to compute the groundstate energy at each allowed fermionic filling (\cref{eq:vqe_loop});
          \item Use a classical computer to obtain the exchange correlation energy \cref{eq:E_QEXC};
          \item From the exchange correlation energy, compute the exchange correlation potential \cref{eq:QEXC_potential};
          \item Interpolate the exchange correlation potential and energy to obtain continuous representations and store on a classical computer.
          \item Use exchange correlation potential and energy in the DFT loop (\cref{algo:DFT_loop}) for any inhomogeneous fermionic Hamiltonian of the form \cref{eq:DFT_gen}.
          \end{enumerate}
          \caption{Obtaining the quantum-enhanced XC potential and energy}
      \end{algorithm}}
      \caption{The algorithmic procedure to generate a classical representation of a quantum-enhanced XC potential and energy}
      \label{algo:XC_potential}
\end{table}

\section{QEDFT examples} \label{sec:examples}

In this section, we present some examples for how the QEDFT algorithm can be applied to the Fermi-Hubbard model.

\subsection{The Fermi-Hubbard dimer}\label{sec:dimer}

Here we will describe each of the necessary steps required for
implementing the QEDFT algorithm for the prototypical $1\times2$ inhomogeneous Fermi-Hubbard
model at half-filling. The model that QEDFT solves is:
\begin{align}
  H_{IH} &= -t(\hat{c}_{1\uparrow}^{\dagger} \hat{c}_{2 \uparrow} + \hat{c}_{2\uparrow}^{\dagger} \hat{c}_{1 \uparrow} + \hat{c}_{1\downarrow}^{\dagger} \hat{c}_{2 \downarrow} + \hat{c}_{2\downarrow}^{\dagger} \hat{c}_{1 \downarrow}) \nonumber \\
  & + U(\hat{n}_{1\uparrow}\hat{n}_{1\downarrow} + \hat{n}_{2\uparrow}\hat{n}_{2\downarrow}) \nonumber \\
  & + v_1(\hat{n}_{1 \uparrow} + \hat{n}_{1 \downarrow})+  v_2(\hat{n}_{2 \uparrow} + \hat{n}_{2 \downarrow}),
  \label{eq:dimer_inhom}
\end{align}
where $U$ is the interaction strength, $t$ is the hopping amplitude, and
$\{v_1,v_2\} $ are the chemical potentials for the different sites that
constitute the external potential. This form of the inhomogeneous Hamiltonian corresponds to
that of \cref{eq:DFT_gen} for two sites. The homogeneous Hamiltonian that is required by the QEDFT
algorithm (\cref{algo:XC_potential}) is equivalent to \cref{eq:dimer_inhom}
without the external potential, which is given by,
\begin{align}
  H_{H} &= -t(\hat{c}_{1\uparrow}^{\dagger} \hat{c}_{2 \uparrow} + \hat{c}_{2\uparrow}^{\dagger} \hat{c}_{1 \uparrow} + \hat{c}_{1\downarrow}^{\dagger} \hat{c}_{2 \downarrow} + \hat{c}_{2\downarrow}^{\dagger} \hat{c}_{1 \downarrow}) \nonumber \\
  & + U(\hat{n}_{1\uparrow}\hat{n}_{1\downarrow} + \hat{n}_{2\uparrow}\hat{n}_{2\downarrow}).
  \label{eq:dimer_hom}
\end{align}

Following the algorithmic protocol presented in \cref{algo:XC_potential}, the
Fermi-Hubbard dimer is first transformed from a complex fermionic representation to
qubit representation via a fermionic encoding. Using the Jordan-Wigner encoding, the terms in the Hamiltonian are transformed as follows:
\begin{equation}
  \label{eq:jordan-wigner}
  \hat{c}^\dagger_i \hat{c}_j + \hat{c}^\dagger_j \hat{c}_i \mapsto \frac{1}{2}(X_i X_j + Y_i Y_j)Z_{i+1} \cdots Z_{j-1}, \,\,\, \hat{n}_i\hat{n}_j \mapsto \frac{1}{4}(I - Z_i)(I - Z_j),
\end{equation}
assuming each spin-orbital has been assigned a unique index with a chosen Jordan-Wigner ordering (e.g. all up orbitals, followed by all down with the lattice sites in a snake ordering~\cite{cade2020strategies}). In the specific case of the half-filled dimer, it is possible to reduce the 4-qubit representation of the Fermi-Hubbard Hamiltonian into a 2-qubit one as only four amplitudes in the statevector are non-zero. The compressed representation is given by~\cite{montanaro2020compressed},
\begin{equation}
  H_{C} = -t ( X \otimes I + I \otimes X) + \frac{U}{2}(I + Z \otimes Z).
\end{equation}
It has been shown in numerical simulations that one layer of the Hamiltonian variational ansatz~\cite{HVA} is required to find the groundstate of this model using VQE~\cite{montanaro2020compressed, cade2020strategies}. Therefore, the state that we prepare using VQE is the following:
\begin{equation} \label{eq:hv_ansatz}
  \ket{\psi(\theta, \phi)} = e^{i\theta ( X \otimes I + I \otimes X) } e^{i \phi (I + Z \otimes Z)} \ket{\psi_0},
\end{equation}
where $\theta, \phi$ are variational parameters, and $\ket{\psi_0} = \ket{-}\otimes\ket{-}$ is the groundstate of $H_{C}$ with $U=0$.

For each fermionic filling the groundstate energy at a fixed $U/t$ needs to be computed, i.e $E(N_e, U/t)$ $\forall N_e \in  \{ 0, 1, 2, 3, 4\}$. For $N_e = 2$, VQE with the ansatz in~\cref{eq:hv_ansatz} can be used. For $N_e = 0, 4$, the number sector is one dimensional and so a quantum computer does not need to be used. The energy for $N_e = 1$ can also be calculated classically using the observation that there are no interacting $U$ terms which reduces the problem to one with hopping terms only. A similar argument can be used in the symmetric case of $N_e=3$.
The calculation of $E(N_e, U/t)$ allows then for the local XC energy to be calculated,
\begin{equation}
  \epsilon_{QEDFT-XC}\left(\frac{N_e}{2}, U/t\right) = \epsilon_{QEDFT}\left(\frac{N_e}{2}, U/t\right) - \epsilon_{MF}\left(\frac{N_e}{2}, U/t\right)  
  \label{eq:xc_energy}
\end{equation}
where the QEDFT energy is found by obtaining the following energies on a quantum machine,
\begin{equation}
  \epsilon_{QEDFT}\left(\frac{N_e}{2}, U/t\right) = \frac{E(N_e/2, U/t)}{4}\quad \forall N_e \in  \{ 0, 1, 2, 3, 4\}.
\end{equation}

The mean field energy can be computed classically by solving the non-interacting homogeneous Fermi-Hubbard model and computing:
\begin{align}
  \epsilon_{MF}\left( \frac{N_e}{2}, U/t \right) &= \frac{U}{2} (\langle \hat{n}_1 \rangle^2 + \langle \hat{n}_2 \rangle^2) \nonumber \\ 
  &-t \langle (\hat{c}_{1\uparrow}^{\dagger} \hat{c}_{2 \uparrow} + \hat{c}_{2\uparrow}^{\dagger} \hat{c}_{1 \uparrow} + \hat{c}_{1\downarrow}^{\dagger} \hat{c}_{2 \downarrow} + \hat{c}_{2\downarrow}^{\dagger} \hat{c}_{1 \downarrow}) \rangle. 
\end{align}
Then, the QEDFT XC potential is found by taking the numerical derivative of \cref{eq:xc_energy} with respect to the density, as shown in~\cref{eq:QEXC_potential}.

Finally, the QEDFT XC potential is interpolated so that it can be queried at any
density between $0$ and $2$. This concludes the first phase of applying QEDFT to
the Fermi-Hubbard dimer which creates a classical queryable function that can be
used in a DFT computation.

Now, having a continuous representation for the XC
energy and potential we can return to solving \cref{eq:dimer_inhom} with lattice
DFT according to the algorithm protocol of \cref{algo:DFT_loop}. Firstly, the
density must be initialised and a common choice is to make it such that it is
proportional to the external potential,
\begin{equation}
  n_{\text{init}} \propto V_{\text{ext}} = ( v_1, v_2).
  \label{eq:init_density}
\end{equation}

The KS potential is then initialised for the two lattice sites,
\begin{align}
  V_1^{KS}(n_1) &= v_1 + \frac{U}{2}n_1 + V_{QEDFT-XC}(n_1) \nonumber \\
  V_2^{KS}(n_2) &= v_2 + \frac{U}{2}n_2 + V_{QEDFT-XC}(n_2),
\end{align}
which allows for the full single particle KS Hamiltonian of the system to
be expressed as,
\begin{align}
  H_{KS}  = & -t(\hat{c}_{1\uparrow}^{\dagger} \hat{c}_{2 \uparrow} + \hat{c}_{2\uparrow}^{\dagger} \hat{c}_{1 \uparrow} + \hat{c}_{1\downarrow}^{\dagger} \hat{c}_{2 \downarrow} + \hat{c}_{2\downarrow}^{\dagger} \hat{c}_{1 \downarrow}) \nonumber \\
  & + V_{1}^{KS}(n_{1 \uparrow}) \hat{n}_{1\uparrow} + V_{1}^{KS}(n_{1 \downarrow}) \hat{n}_{1\downarrow} \nonumber \\
  & + V_{2}^{KS}(n_{2 \uparrow}) \hat{n}_{2\uparrow} + V_{2}^{KS}(n_{2 \downarrow}) \hat{n}_{2\downarrow},
\end{align}
where we have separated the spin channels, which we consider splitting into two independent problems as
we deal with the spin unpolarised case by dealing with the total density at site $i$ as $n_{i} = n_{i \uparrow} + n_{i \downarrow}$.
The KS system of equations is then diagonalised, i.e $ H_{KS} \phi_{j} = \epsilon_{j} \phi_{j} $, where $\epsilon_{j}$ are the KS eigenvalues and
$\phi_{j}$ are the KS eigenvectors.
Explicitly, for the Fermi-Hubbard dimer, the KS matrix for a given spin channel is,
defined by,
\begin{equation}
  H_{KS} =
  \begin{pmatrix}
    V_1^{KS}(n_1) & -t \\
    -t & V_1^{KS}(n_2)
  \end{pmatrix}.
  \label{eq:KS_matrix}
\end{equation}

The spectrum of the KS problem consists of the
eigenvalues $ \{\epsilon_1,\epsilon_2 \}$ and eigenvectors $\{ \phi_{1} , \phi_{2} \}$, that
are of dimension $L$. From the KS eigenvectors a new density is constructed, for the half-filled system this is:
\begin{align}
n_1 & = 2|\phi_1(1)|^2,   \nonumber \\
n_2 & = 2|\phi_1(2)|^2,
\end{align}
where $\phi_i(j)$ is the $j^\text{th}$ element of the $i^\text{th}$ KS eigenvector.
This density is then used as the input to the next iteration of the DFT iterative
algorithm, i.e. \cref{eq:init_density} is replaced with these values. At each iteration
the energy of the parent Hamiltonian, i.e. \cref{eq:dimer_inhom}, is calculated using:
\begin{align}
  E_{GS}(\mathbf{n}^{GS}, U) &= 2\epsilon_{1} -V_{QEDFT-XC}(n_1, U)n_1 - V_{QEDFT-XC}(n_2, U)n_2 \nonumber \\
  & - \frac{1}{4}U(n_{1}^2 + n_{2}^2) + \epsilon_{QEDFT-XC}(n_1,U) + \epsilon_{QEDFT-XC}(n_2,U).
\end{align}

The algorithm concludes when the difference between the energy at two successive
iterations is less than a chosen tolerance $\delta$. At this point the results
of the DFT algorithm for a chosen functional can be compared to the results
produced via an exact method, which is possible as the Fermi-Hubbard dimer can
be efficiently solved both analytically and numerically due to its small size.

\subsection{More general Fermi-Hubbard models}\label{sec:beyond-dimer}

We extend the dimer by adding more sites for a specified geometry to produce
the generalised inhomogeneous Fermi-Hubbard model given by,
\begin{equation} H = -t \sum_{\langle ij \rangle, \sigma} (\hat{c}_{i
\sigma}^{\dagger}\hat{c}_{j \sigma} + \hat{c}_{j \sigma}^{\dagger}\hat{c}_{i \sigma}) +
\sum_{i}^{L} U \hat{n}_{i \uparrow} \hat{n}_{i \downarrow} + \sum_{i}^{L} v_i^{ext}
\hat{n}_i
\label{eq:IHM}
\end{equation}
where $v_i^{ext}$ is the external lattice potential, $U$ is the interaction
strength, $t$ is the hopping parameter and $L$ is the system size.
Here $\langle ij\rangle$ indicates a sum over nearest neighbours, and $\sigma$ denotes the spin.
KS theory then maps \cref{eq:IHM} into the effective KS system,
where the many body interaction term is absorbed into a one-body potential
$V_{i}^{eff}$,
\begin{equation} H_{KS, \sigma}(n_i) = -t \sum_{\langle ij \rangle}
(\hat{c}_{i\sigma}^{\dagger}\hat{c}_{j\sigma} + \hat{c}_{j\sigma}^{\dagger}\hat{c}_{i\sigma}) + \sum_{i}^L V_{i \sigma}^{eff}(n_{i \sigma})
\hat{c}_{i \sigma}^{\dagger}\hat{c}_{i \sigma},
\label{eq:KS_IFH}
\end{equation}
for a given spin channel $\sigma$, which we now omit as we treat both spin channels
equally going forward by setting the density variable as $n_i = n_{i \uparrow} + n_{i \downarrow}$. The effective KS potential for the generalised Fermi-Hubbard model
is written as a combination of the external, Hartree, and XC potentials,
\begin{equation}  V_{i}^{eff}(n_i) = v_{i}^{ext} + \frac{U}{2} n_i
+ V_{QE-XC}(n_i).
\end{equation}
$V_{QE-XC}(n_i)$ is the unknown XC potential, which is approximated according to
the methods prescribed for \cref{eq:QEXC_potential} described in
\cref{algo:XC_potential}. Specifically, the XC energy for the Fermi-Hubbard model is,
\begin{equation}
  E_{QE-XC}(n_i, U) = \frac{1}{L} E_{QE-hom}(n_i, U) - \frac{t}{L} \sum_{\langle ij\rangle\sigma} \langle \hat{c}_{i\sigma}^{\dagger}\hat{c}_{j\sigma} + \hat{c}_{j\sigma}^{\dagger} \hat{c}_{i\sigma} \rangle - \frac{1}{L} \sum_{i}^{L}\frac{U}{4} \langle n_{i}^{2} \rangle
  \label{eq:FH_XC_energy}
\end{equation}
 where $E_{QE-hom}(n_i, U)$ is the energy of the homogeneous Fermi-Hubbard model,
retrieved from \cref{eq:IHM} by setting $v_{i}^{ext}=0$, and computed using VQE
of a specified depth and occupation number. In the local density approximation
the total homogeneous energy for the Fermi-Hubbard model is defined as the sum
of the average energy in a homogeneous Fermi-Hubbard over the entire chain. Once
\cref{eq:FH_XC_energy} is computed and an ansatz for the initial density is made, the XC potential can be obtained by
\cref{eq:QEXC_potential}. This potential determines the KS system of equations to be solved. For example, for the 1D Hubbard chain with open boundary conditions, the KS matrix is
\begin{equation}
  H_{\text{KS}} =
  \begin{pmatrix}
    V_1^{\text{eff}}(n_1) & -t  &          &             &              & 0 \\
    -t  & V_2^{\text{eff}}(n_2) & -t  &             &              & \\
    & -t  & V_3^{\text{eff}}(n_3) & \ddots      &              & \\
    &          & \ddots   & \ddots      & -t  & \\
    &          &          & -t & V_{L-1}^{\text{eff}}(n_{L-1}) & -t \\
    0        &          &          &             & -t      & V_L^{\text{eff}}(n_L) \\
  \end{pmatrix},
  \label{eq:KS_matrix_IFH}
\end{equation}

\begin{figure}[!t] \centering
  \includegraphics[width=\textwidth]{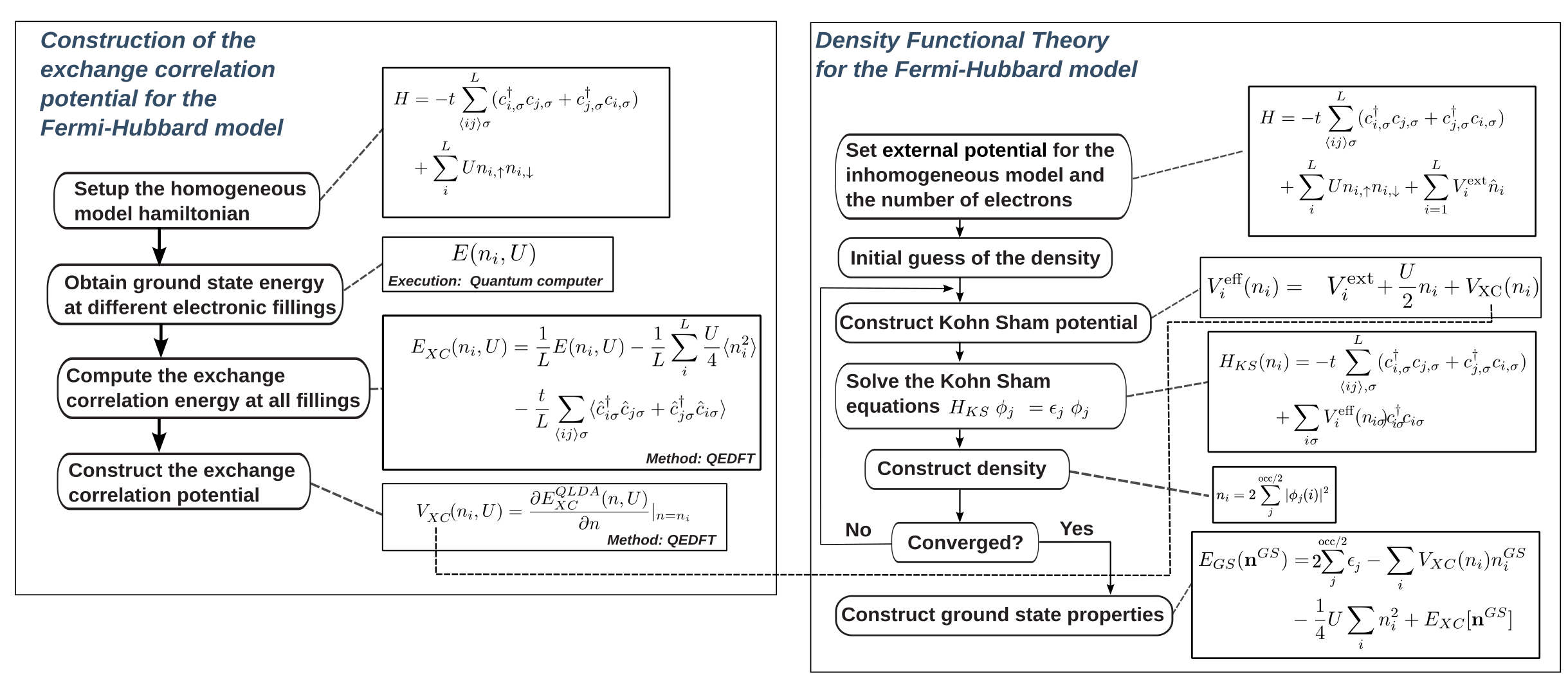}
  \caption{The general flow diagram of how the quantum and classical
components interact in the QEDFT algorithm for the Fermi-Hubbard model. On the left
hand side we illustrate how the exchange correlation potential is constructed
within QEDFT. On the right hand side we show how the
DFT loop is implemented within this hybrid scheme. We have highlighted by
``Execution: Quantum Computer`` the computations which are run on the quantum
computer, while the rest of the computations are all run on a classical
computer.}
\label{fig:algo_design}
\end{figure}

The result of this diagonalisation
are the KS eigenvalues, $\{ \epsilon_1, \hdots, \epsilon_L \}$, and the KS
eigenvectors, $\{ \phi_{1}, \hdots, \phi_{L} \}$. The
KS eigenvectors are used to construct a new approximation to the density via
\cref{eq:KS_density}, which is then mixed with the previous density
\cref{eq:mixing} for numerical stability, which can be then used to construct a
new KS matrix \cref{eq:KS_matrix_IFH}. At convergence the DFT approximation to
the groundstate energy is,
\begin{align}
  E_{GS}(\mathbf{n}^{GS}) &= 2\sum_{i\leq{\rm occ}/2} \epsilon_{i}^{KS} - \sum_{i\in\mathcal{L}}V^{XC}(n_i)n_i^{GS} \nonumber \\ 
  &- \frac{1}{4}U\sum_{i\in\mathcal{L}}n_{i}^2 + \sum_{i\in\mathcal{L}} E_{QE-XC}(n_i,U),
\end{align}
where only the first $N_{e}$ KS eigenvalues are included in the summation, with
$N_e$ corresponding to the fermion filling of the inhomogeneous model \cref{eq:IHM}.
The entire protocol for the generalised Fermi-Hubbard system is illustrated in
\cref{fig:algo_design}.



\section{Results}\label{sec:numerics}

To test and validate the QEDFT algorithm described in \cref{sec:QEDFT}, we performed numerical simulations on classical hardware for 1D and 2D Fermi-Hubbard systems, and computations for 1D Fermi-Hubbard systems using data from Google's Rainbow quantum hardware~\cite{stanisic2022observing}.
We present these results in the following sections, followed by numerical results and analysis of the XC potentials produced by QEDFT and conventional DFT methods that gives insight into the performance of the QEDFT algorithm.
The computational details used to generate these results are discussed in detail in \cref{app:comp_details}.


\subsection{1D Fermi-Hubbard results}
In this section, we present results of the QEDFT algorithm for 1D Fermi-Hubbard systems, both in simulations and in computations using data from real quantum hardware, and comparing these to conventional DFT methods.
1D Fermi-Hubbard systems are readily solvable classically, both by analytic and numerical methods.
The purpose of testing the algorithm on these systems is not, therefore, to compare the QEDFT performance to the best known classical methods, but rather to test the algorithm on well-characterised systems and assess (1)~how close it gets to the correct solution, and (2)~to compare its performance with classical DFT methods that scale to 2D and other systems (but which are not the optimal methods for 1D systems).

To obtain (good approximations) to the exact 1D solutions against which to benchmark, we use exact diagonalisation (ED) for system sizes small enough for this to be possible, and classical DFT with the BALDA functional~\cite{xianlong2006bethe} for system sizes beyond ED.
The BALDA functional is based on exact solvability of the 1D Fermi Hubbard model in the thermodynamic limit.
So although BALDA DFT is a classical DFT method, for finite 1D Fermi Hubbard systems it is known to give results that are close to the exact solution.

For the 1D simulations, we take Hartree-Fock as a representative classical DFT method to compare and contrast QEDFT to.
This is of course an artificial comparison, as 1D systems can be solved very effectively classical by other DFT methods such as BALDA DFT.
But as QEDFT is a general method that is not tailored to 1D systems, it is useful to compare also to conventional DFT methods that are similarly generally applicable.
Thus the comparison to Hartree-Fock DFT (HF) on 1D Fermi Hubbard systems allows us to gain insight into how QEDFT compares to conventional DFT methods that are applicable to general systems, but on a simple, well-characterised solvable model where we can also compare both to exact solutions.

Because the results in this section are obtained by simulating QEDFT (or more precisely, the VQE subroutine) on a classical computer, we can only compute the VQE functionals used in QEDFT on small lattice sizes.
For such small lattices, simulated VQE with perfect quantum gates can readily find the exact ground state, so that QEDFT reduces to DFT with ED functionals.
Therefore, in order to gain insight into how QEDFT performs when the VQE functionals are \emph{not} perfect, which will always be the case when one runs VQE on real quantum hardware with lattice sizes that are beyond the reach of classical simulation, we deliberately hobble the VQE by restricting the VQE circuit depth such that it is not able to obtain the exact ground state.

\subsubsection{Numerical simulation results}
\begin{figure}[t!] \centering
  \includegraphics[width=0.8\linewidth]{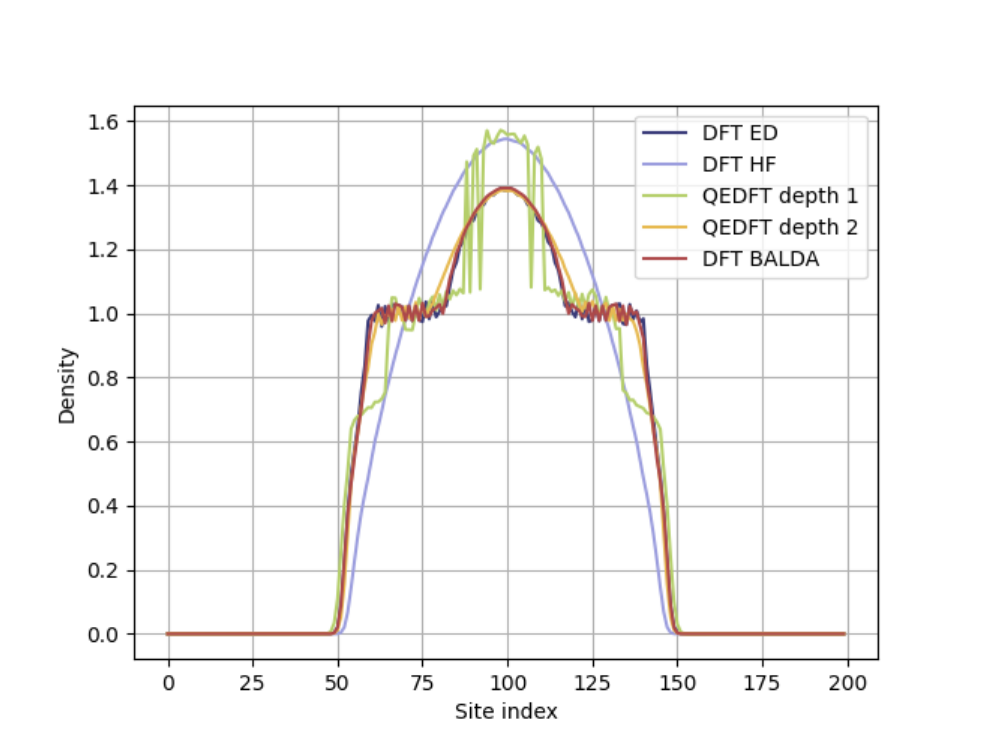}
  \caption{The ground state density of the $1 \times 200$ Fermi Hubbard model for $U/t=10$ and at quarter filling within a confining quadratic potential using a QEDFT functional generated on a $1 \times 12$ homogeneous Fermi-Hubbard system.
    The presence of Mott plateaus, characteristic of correlations in the system, are captured by the shallow depth VQE functionals, albeit less so at depth 1.
    These features are absent in the Hartree Fock approach.
  }
\label{fig:fig6}
\end{figure}

In \cref{fig:fig6} we show the results of applying QEDFT for a $1\times200$ instance of the Fermi Hubbard model using a functional generated by the emulation of VQE applied to a $1\times12$ Hubbard chain.
In these simulations we use the BALDA functional as the exact benchmark, as performing ED simulations for these system sizes is not possible.
The system parameters are $U/t=10$ at quarter-filling ($N_e=100$) within a confining quadratic potential $v_{i}^{\text{ext}} = (i-L/2)^2/L$, with $L=200$.
This model is known to exhibit regions of coexisting phases between metallic (compressible) and Mott insulating (incompressible) characteristics~\cite{xianlong2006bethe}.
Such a scenario is illustrated in \cref{fig:fig6} in the BALDA density profile at the centre of the trap, where a plateau emerges and the density is constrained to $n_i=1$.
This plateau is determined by the competition between $v_{i}^{ext} + Un_i/2$ and $V_{XC}$ in the KS potential, which are separated by the XC potential discontinuity at $n_i=1$.

Importantly, the QEDFT methods capture the coexistence of the localised incompressible region with that of the metallic compressible phase, a key feature of the correlated regime.
The correlated driven plateaus are entirely absent from the HF result and hence serve as a demonstration that low depth and low quality QEDFT functionals can be qualitatively superior in their descriptive capabilities in the strongly interacting regime.
Additional analysis of the energy convergence as function of the DFT iterations is presented in~\cref{app:qedft_1d}.
These results give evidence that QEDFT usually achieves better accuracy than Hartree-Fock DFT its own, without requiring extremely accurate quantum computations.

\begin{figure}[t!] \centering
  \includegraphics[width=\linewidth]{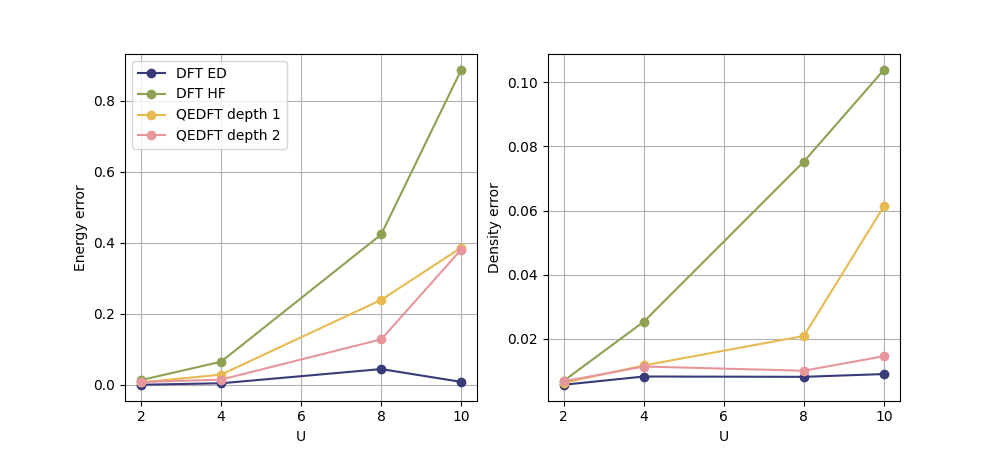}
  \caption{The $1 \times 200$ Fermi Hubbard model at quarter-filling with a quadratic external potential and its respective errors in energy (left panel) and density (right panel) for different interaction strengths.
    All simulations use DFT and the BALDA DFT functional is used as the exact benchmark.
  }
\label{fig:fig8}
\end{figure}

To demonstrate that our emulated results are not valid only for a single $U/t$ value, we present the effect of varying the interaction strength on the $1\times200$ system, shown in \cref{fig:fig8}.
The errors are measured relative to the BALDA DFT method.
Here we find that for all values of $U/t$ examined the QEDFT densities are more accurate than the HF densities, and similarly for the ground state energy.
This shows that for large 1D systems QEDFT based on restrictive quantum emulations achieves better results than HF DFT alone, for a system size currently well beyond the reach of standard quantum VQE.
However, the accuracy deteriorates at larger interaction strengths, in particular for HF and QEDFT depth 1.
This is related to the known issue of sampling near unity in the compressible phase with DFT methods that use functionals with XC discontinuities~\cite{xianlong2006bethe}.
At $U/t=10$ we see for QEDFT depth 1 that there is a jump in the density error, which we associate with the emergence of the Mott plateaus near the centre of the trap.
For lower values of $U/t$, the QEDFT functionals are comparable and competitive with the essentially exact BALDA solution.
Further supporting results are described in~\cref{app:qedft_1d}, in particular relating to smaller 1D systems and the effect of finite size scaling on the functional.

\subsubsection{Results from real quantum hardware data}\label{sec:real_quantum_hardware}
\begin{figure}[t!] \centering
  \includegraphics[width=0.8\linewidth]{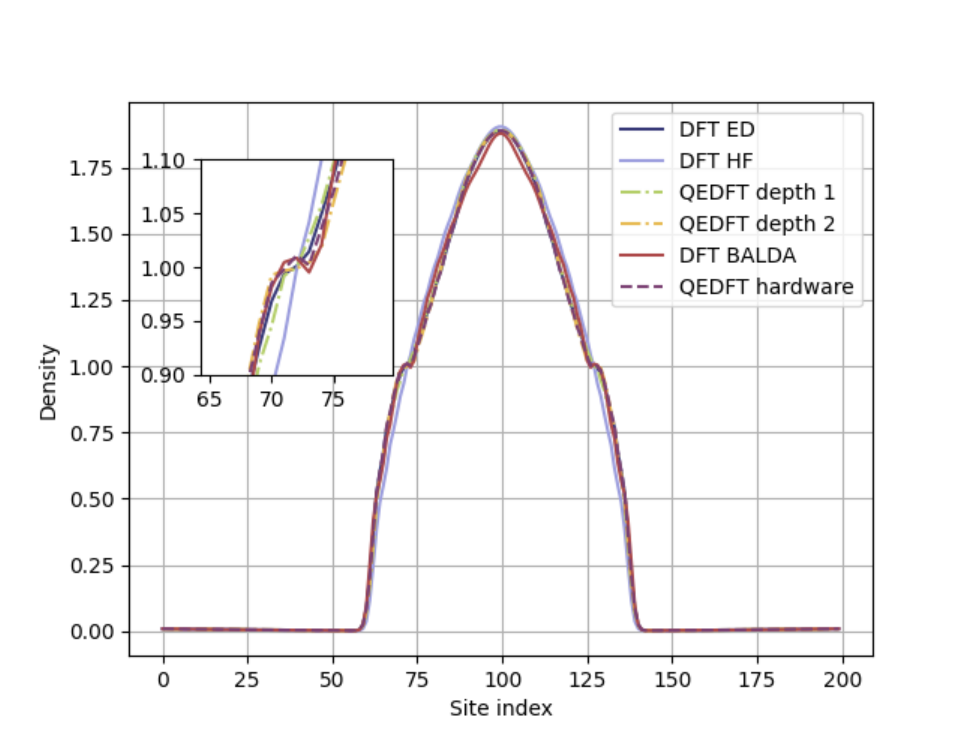}
  \caption{The ground state density of $1 \times 200$ Fermi Hubbard model for $U=4$ at quarter filling within a confining quadratic potential using a QEDFT functional generated on a $1 \times 8$ homogeneous Fermi-Hubbard system from quantum hardware, as well as emulation.
    The inset highlights the region where the Mott plateaus emerge, which is the signature of correlations that are not captured with the HF approach.
  }
\label{fig:fig7}
\end{figure}

We also apply the $1\times8$ QEDFT functional derived from data taken on real quantum hardware, to the much larger $1\times200$ inhomogeneous Fermi-Hubbard model in the presence of an external quadratic potential at $U/t=4$.
Our parameter selection here was constrained by the parameter selections used for the hardware study~\cite{stanisic2022observing}.
We see in \cref{fig:fig7} that the QEDFT hardware also captures the presence of Mott plateaus near $n=1$, albeit less pronounced than our emulated results as in this case $U/t=4$ instead of $U/t=10$.
Note that this feature is completely absent in the HF solution.
These results give evidence that QEDFT is still able to give good solutions even in the presence of noise and errors that are inherent to running on real, pre-fault-tolerant quantum hardware.

\subsection{2D Fermi-Hubbard results}

\begin{figure}[t!] \centering
  \includegraphics[scale=0.8]{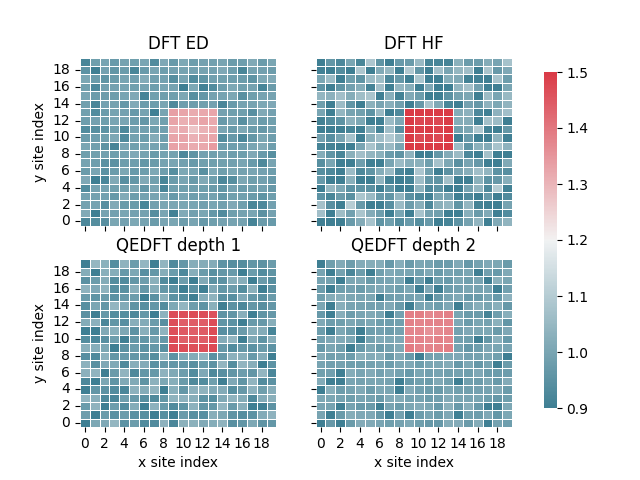}
  \caption{The density per site of the $20 \times 20$ Fermi Hubbard model at half-filling at $U/t=4$ based on DFT calculations using functionals generated on a $3 \times 3$ lattice.
    The external potential is an impurity potential that concentrates fermions to a square at the centre of the lattice. As this model cannot be solved exactly, the DFT ED solution acts as an approximation to the ground state. 
    }
\label{fig:fig9}
\end{figure}

Applying the functional generated in a small scale instance to an inhomogeneous larger system serves as a first step to make quantum-generated data on small devices useful for relevant applications.
In order to test the validity of our results, we treat more challenging systems, where access to an exact functional is not possible.
In this section, we present results of applying QEDFT to 2D Fermi Hubbard systems, using classical simulations and computations using real hardware data, as well as computations using classical DFT methods.

We do this by extending our analysis to a $20\times20$ inhomogeneous Fermi-Hubbard system for $U/t=4$ at quarter-filling.
The inhomogeneities are generated by a cluster of impurites at the centre of the lattice in the presence of a weak background disordered potential.
Exact solutions for this system would require diagonalising a matrix of dimensions $4^{20\times20}\times4^{20\times20}$, which is not possible with classical computers.
DFT, on the other hand, only requires the diagonalistion of a matrix with dimensions $(20\times20)\times(20\times20)$ which can be easily achieved.
\cref{fig:fig9} shows the DFT densities for the $20\times20$ system using the $3\times3$ QEDFT XC functional (see~\cref{app:qedft_2d}).
It is clear that all of the DFT methods are able to distinguish where the impurities are located.
The absence of an exact method for large 2D systems (unlike 1D systems where the BALDA functional exists, as well as other accurate 1D methods such as DMRG) means there is no exact ground truth comparison, and we choose to compare against the DFT ED method.

We see that the QEDFT functional of depth 2 is more in likeness to the DFT ED result, both in contrast to the classical HF functional and QEDFT depth 1.
Indeed, these results are comparable to the results we obtained for the densities of the $1\times200$ Fermi-Hubbard system, where QEDFT of depth 1 approximates the HF solution better than the exact solution.
Thus, despite the lack of an exact solution for the ground state density, it is at least suggestive that QEDFT at depth 2 is incorporating effects that are not present in the HF solution.
These results thus lend further support to the evidence that QEDFT, in this case applied to 2D Fermi-Hubbard systems, gives useful results even when only small quantum devices are available and where hardware noise and errors prevent good solutions being obtained by VQE.
Further analysis and results, in particular computations using real hardware data, are given in \cref{app:qedft_2d}, on smaller 2D systems as well as their finite size scaling.



\subsection{Analysis of XC functionals and potentials}
\begin{figure}[!t] \centering
  \includegraphics[width=\linewidth]{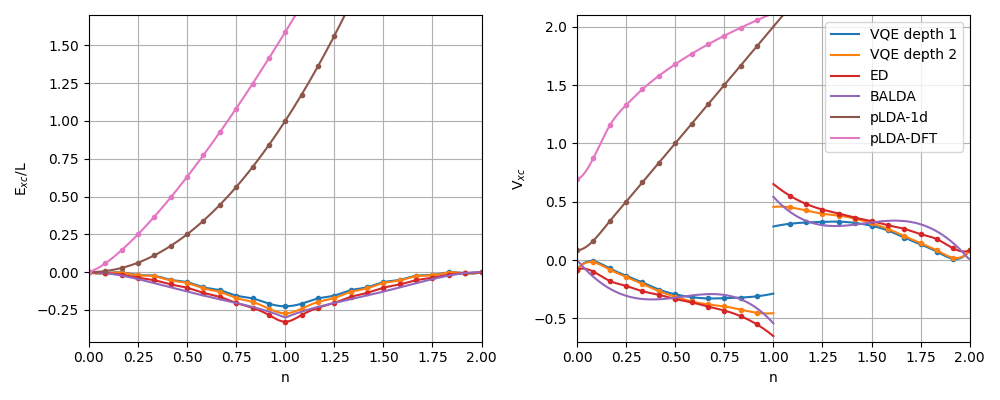}
  \caption{The XC energy (left) and potential (right) for the $1 \times 12$
    Fermi-Hubbard model at $U/t=4$ using different methods. The solid lines are
    produced using a cubic spline. The VQE results are obtained using a
    classical emulator. Note that the continuous LDA approximations perform badly in this system.}
\label{fig:1d_XC_fig_a}
\end{figure}

Using the algorithm presented in \cref{algo:XC_potential}, we generate the exchange correlation energies, from which we differentiate and apply a cubic polynomial spline to obtain the XC potential, as illustrated in~\cref{fig:1d_XC_fig_a}. The purpose of generating these correlation functionals is to query them inside the DFT algorithm, hence the necessity for obtaining a continuous representation as a function of the number of electrons.
To gain insight into how QEDFT and other DFT methods perform, it is therefore valuable to investigate and analyse the functionals that are generated by these different methods.

In order to do so, we must first discuss the general properties of the XC functionals for Fermi-Hubbard systems.
The continuous lines in \cref{fig:1d_XC_fig_a} represent cubic spline fits for each functional, where the XC potential is defined piecewise $V_{XC}^{\text{left}}$ for $n \leq 1$ and $V_{XC}^{\text{right}}$ for $n > 1$.
We note that by replacing the discrete points with a continuous spline we introduce another source of approximate error, in addition and of different nature than the error introduced by approximating the XC functional.
On the other hand, if the continuous representation is constructed on a dense enough mesh, and if the VQE depth is sufficient to within a desired error tolerance of the ED functional, then finite size physical errors, as well as those related to the interpolation scheme, can be reduced.

The XC potential has a discontinuity at $n=1$ due to the cusp in the exchange correlation energy, commonly known as the derivative discontinuity (DD). The DD is a known property
of the exact XC potential which a number of classical approximations fail to
reproduce~\cite{DDI} and has consequences for the description of properties
relating to the fundamental gap of the system. All functionals presented in
\cref{fig:1d_XC_fig_a} exhibit this discontinuity, with the exception of the
``pseudo`` functionals, and the depth 2 VQE functionals approximate this
discontinuity with higher accuracy in regions both near and away from where the
discontinuity occurs compared to depth 1. The ``pseudo-DFT'' functional, $E_{xc}(n) = 2^{-4/3}Un^{4/3}$ is based on the local density approximation applied in
ab-initio methods, specifically for the homogeneous electron gas, and evidently
is not a good approximantion for the Fermi-Hubbard system. Similarly, the
``pseudo-1D'' functional, $E_{xc} = 0.25 U n^2$, originates from the 1D
equivalent. Not only do these functionals omit the DD but they also
qualitatively arrive at values which are not comparable for either the XC energy
or the XC potential. As these pseudo functionals are so erroneous - which we find is
also the case in 2D - we chose not to analyse or
implement them as part of the DFT study. We emphasise that the BALDA
functional is valid in the limit of an infinite chain, but indeed the results
for the $1\times12$ chain approximate this behaviour.

\begin{figure}[!t]
  \includegraphics[width=\linewidth]{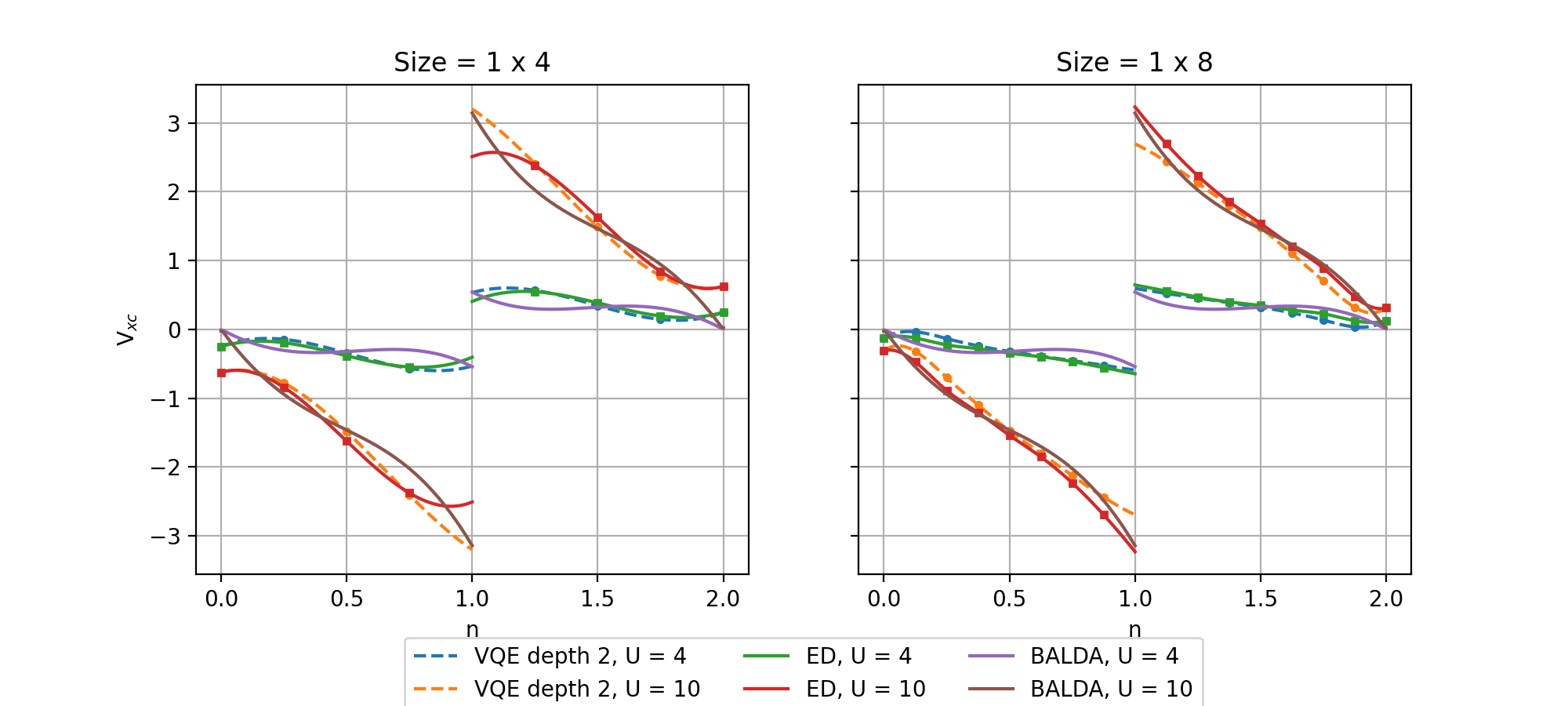}
  \caption{The effect of $U/t$ and system size on the XC potential for 1D Fermi Hubbard models. Note that as $U/t$ increases so to does the derivative discontinuity at $n=1$ and by increasing the system size the mesh upon which the functional is splined becomes denser, reducing finite size errors.  }
\label{fig:fig3}
\end{figure}

Having established that it is possible to generate the XC potential using VQE, in
\cref{fig:fig3} we analyse the effect that different parameters, \emph{i.e} correlation strength and system size, have on the resulting functionals. We observe that VQE functionals
qualitatively agree with the ED and BALDA results, \emph{e.g} capturing the DD and its general shape, but deteriorate quantitatively as the interaction strength increases. However, as the initial VQE state preparation corresponds to the $U=0$ configuration, there are expected inaccuracies of the VQE circuits used for larger interaction strengths. Further detailed analysis that supports these observations is also presented in \cref{app:xc_error}. We see that the continuous representation is constructed on a denser mesh for larger systems, as illustrated via the comparison of the $1\times8$ and $1\times12$ functionals. Thus if the VQE functional is generated on finer meshes for sufficient depth, then finite size physical errors, as well as those related to the interpolation scheme, can be reduced.

\begin{figure}[t!] \centering
  \includegraphics[width=\linewidth]{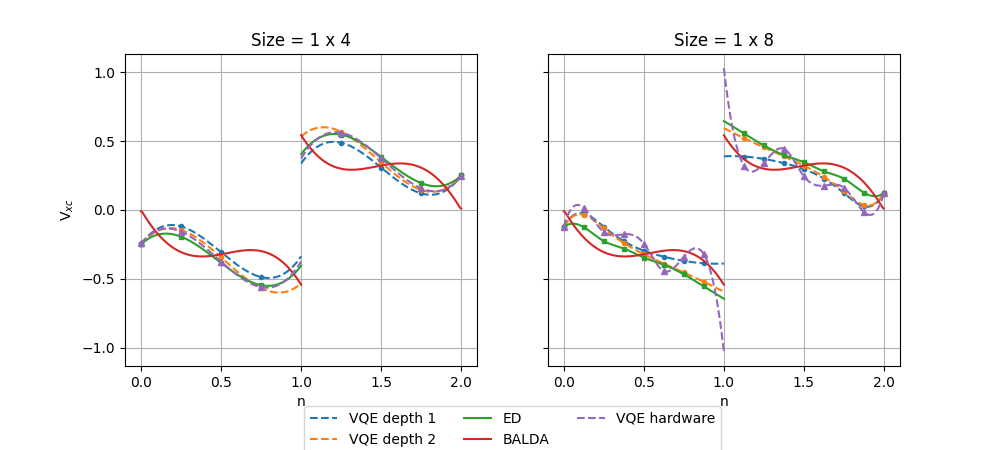}
  \caption{The effect of system size on the XC potential for 1D and 2D Fermi Hubbard models at $U/t=4$ run on quantum hardware and classical simulations. Note that for the $1\times4$ system the quantum hardware functional is competitive with classical ED, while the presence of noise becomes apparent in the larger $1\times8$ hardware results. }
\label{fig:fig4}
\end{figure}

In \cref{fig:fig4} we show the VQE functional properties at fixed $U/t=4$ across different system sizes for real quantum hardware data, details of which are specified in \cref{app:comp_details}. The $1\times4$ system shows the most accurate results, where the hardware results consistently fall between the simulated VQE depth 2 and DFT ED XC potentials. However, as the system size is doubled to $1\times8$ there is the
manifestation of noise in the hardware functional, which has effects on both
the raw data as well as creating oscillations in the subsequent interpolation, in particular in the density region near 1, so much so that it artificially enhances the DD by a factor of 2. Therefore, while we have shown that it is possible to generate competitive XC functionals using real hardware, the quality of the functionals worsens as the system size increases, due to noise in the measurements.

\begin{figure}[t!] \centering
  \includegraphics[width=\linewidth]{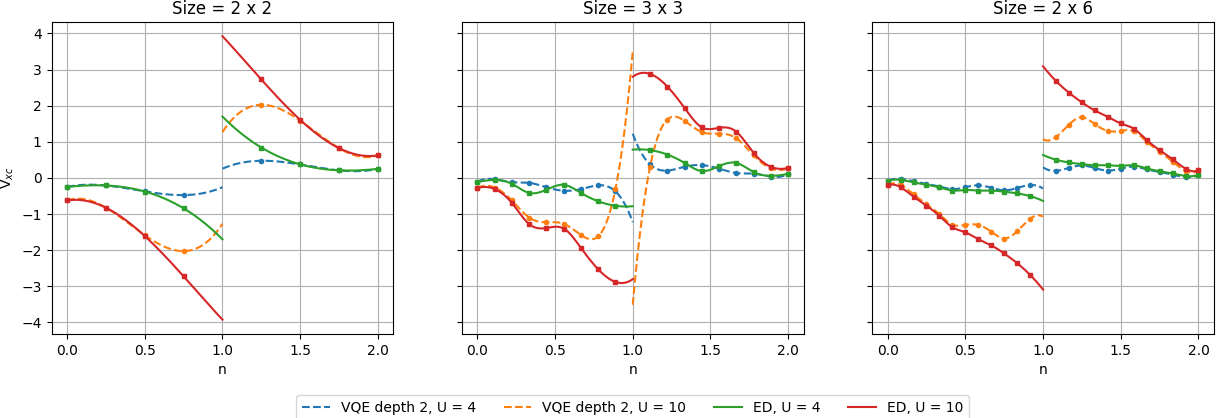}
  \caption{The effect of $U$ and system size on the XC potential for 2D Fermi Hubbard models. The accuracy of the 2D VQE XC potentials at depth 2 is worse when compared to their 1D counterparts. In particular, for low interaction strengths their accuracy is competitive with classical ED, but there are larger errors for stronger interactions. }
\label{fig:fig5}
\end{figure}

We furthermore extend the analysis of the effect of system size and interaction strength to VQE functionals generated on 2D lattices, shown in \cref{fig:fig5}. As the Bethe ansatz
applies only in 1D, we do not compare to the BALDA functional in 2D. We highlight that the curvature and DD of the functional in 2D attains a different quality and is visibly different in character for the configurations studied, e.g the $3 \times 3$ functional exhibits steps near $n=1.5$ and $n=2$, while $2 \times 6$ is mostly monotonic. There is a degradation of the functional near half-filling at $U/t=10$ which is present in each 2D emulated functional that impacts the continuous representation, particularly noticable for the $3\times3$ system. However, there is considerably better agreement at $U/t=4$, indicating that low-depth functionals can approximate the ED functionals for lower values of the interaction strength in 2D, with additional evidence explored in \cref{app:xc_error}.

\section{Conclusions}\label{sec:conclusions}

Density Functional Theory is the de facto standard for simulating molecular and
materials groundstate properties on traditional computers. Its overwhelming
popularity and success relies on choosing appropriate classical approximations to
the unknown exchange correlation energy functional on a system-by-system basis.
For many cases, especially those with strongly correlated electrons, the
classical approximations are inadequate, and sophisticated approaches beyond DFT
are needed that are not efficiently scalable on traditional computers. Simulating
materials is regarded as a promising application for quantum computers
\cite{clinton2022towards} and has so far largely been viewed through the lens of
hybrid quantum/classical algorithms to make the most of near-term noisy quantum
resources. In this work we have introduced a new hybrid quantum/classical
approach called \emph{quantum-enhanced DFT} (QEDFT) that combines DFT
with quantum computation. We have presented this formalism in generality within
the specific framework of lattice DFT and applied it to various instances of the
1D and 2D Fermi-Hubbard models. Low depth VQE simulations and hardware experiments
are used to construct approximations to the XC energy functional, which are then
available to be used as part of the DFT algorithm.

In the majority of cases considered, QEDFT outperforms pure quantum VQE alone,
with respect to the accuracy of the groundstate energy and density of both the
1D and 2D inhomogeneous Fermi-Hubbard model. It also provides ground state properties which
are more accurate than classical Hartree Fock theory. QEDFT captures exact
properties of the XC potential functional, such as the derivative discontinuity
at half-filling, which are not present in many well-known classical
approximations. The derivative discontinuity and groundstate properties from
QEDFT show impressive robustness to hardware noise and shallow variational
layers. 
A hallmark of the QEDFT method is that it does not rely on highly accurate quantum computations to produce groundstate properties
of either inhomogeneous or homogeneous lattice systems. Another quintessential
feature is that QEDFT can be applied to large many-body condensed matter systems
using only small quantum hardware. We achieve this by generating continuous
representations of the QEDFT XC potential using classical interpolation schemes.
In particular, we have shown this by solving $1\times200$ and $20\times20$
Fermi-Hubbard models using QEDFT, demonstrating that shallow VQE functionals of
at most two variational layers can unambiguously capture Mott plateaus in the
groundstate densities of large inhomogeneous models. We have also shown that our results improve by increasing the system size that the XC potential is generated from, highlighting that our approach can benefit from running quantum simulations on larger devices.  


Extending the application of QEDFT beyond Fermi-Hubbard models to realistic
atomic systems requires broadening the scope of the inhomogeneous Hamiltonians under
consideration. In particular, the inclusion of additional effects is
imperative to enhance the descriptive capabilities of QEDFT functionals to
capture a rich variety of important physical processes such as multiorbital couplings,
beyond nearest-neighbour interactions, lattice connectivity, and electron-phonon
interactions, to name only a few important effects.
Another important consideration
for realistic systems is the effect of the underlying basis set used to
represent continuous inhomogeneous models. This choice is well known to have
significant technical consequences for how to setup the DFT problem and has
recently also been explored in the context of quantum
simulation~\cite{clinton2022towards, PhysRevResearch.5.013200}. The basis set
will also be an important consideration when building the QEDFT XC functionals
that are used for real atomic systems, as the optimal representation of the
homogeneous problem is needed to construct it. Additionally,
including non-local effects from the density into the QEDFT functional, in the
spirit of the classical GGA~\cite{GGA} approximation, presents complementary
ways in which the QEDFT method can be improved upon when applied more generally.

In conclusion, QEDFT presents a new way in which to bring together DFT
with quantum simulation, that is applicable across a broad range of physical
fermionic Hamiltonians describing real atomic, molecular, and materials
systems. It has the benefits of being noise resilient and only requires modest
quantum resources on small quantum machines that can produce accurate
groundstate properties of systems not able be simulated directly. We have
implemented QEDFT for the prototypical Fermi-Hubbard lattice model and
demonstrated that it is a powerful hybrid quantum/classical approach that can
compete with state-of-the-art classical methods and outperform conventional
variational quantum algorithms. QEDFT holds the potential to optimise the
utility of near-term quantum machines for accurate simulations of many-body
quantum systems such as materials and molecules.

\section*{Acknowledgements}

We would like to thank the rest of the Phasecraft team for many helpful and insightful discussions. This work was supported by the Innovate UK (grant no. 44167).


\appendix
\section*{Appendix}

\section{Computational methods}\label{app:comp_details}


For the VQE simulations, we implemented the Hamiltonian variational
ansatz~\cite{HVA} which has been shown to be promising for solving the
Fermi-Hubbard model~\cite{cade2020strategies}. The starting state was taken
to be the groundstate of the $U=0$ Fermi-Hubbard model which can be prepared
efficiently on a quantum computer using Givens rotations~\cite{Zhang2018} and
the commuting sets were auto-generated using a greedy graph colouring
algorithm. The quantum circuit simulations for VQE were run using the Julia
package Yao~\cite{yao}. Since all the simulations generated the exact VQE
state, i.e. no measurements or noise, we were able to use Yao's automatic
differentiation framework to speed up calculations of the gradient to be used
with the L-BFGS~\cite{LBFGS} optimiser in NLopt~\cite{NLopt}.

To analyse the performance of QEDFT on quantum hardware, we were able to make use of previous Fermi-Hubbard VQE experiments that had been carried out on Google's 23-qubit Rainbow chip \cite{stanisic2022observing}. The three instances of the model that were considered in the paper and that we consider here are: the $1\times4$ Fermi-Hubbard model with VQE depth 2; the $1\times8$ Fermi-Hubbard model with depth 1; and the $2\times4$ Fermi-Hubbard model with
depth 1. For each of these models we generated a QEDFT XC potential, referred to as the QEDFT hardware functional, which is then used inside a DFT
loop for the original homogeneous system and complementary inhomogeneous
systems. Note that the ansatz that was run on the hardware is not identical to the ansatz for the exact simulations, which is why the hardware may sometimes produce different results to simulated VQE, up to noise.

The lattice DFT algorithm is implemented according to~\cref{algo:DFT_loop} and the self 
consistent protocol is run, unless otherwise specified, for 500 DFT iterations using a density mixing of $\alpha=0.95$ with the initial density ansatz being chosen such that it is proportional to the external potential. 

\section{Error analysis of the XC functional}\label{app:xc_error}

\begin{figure}[!t] \centering
  \includegraphics[width=\linewidth]{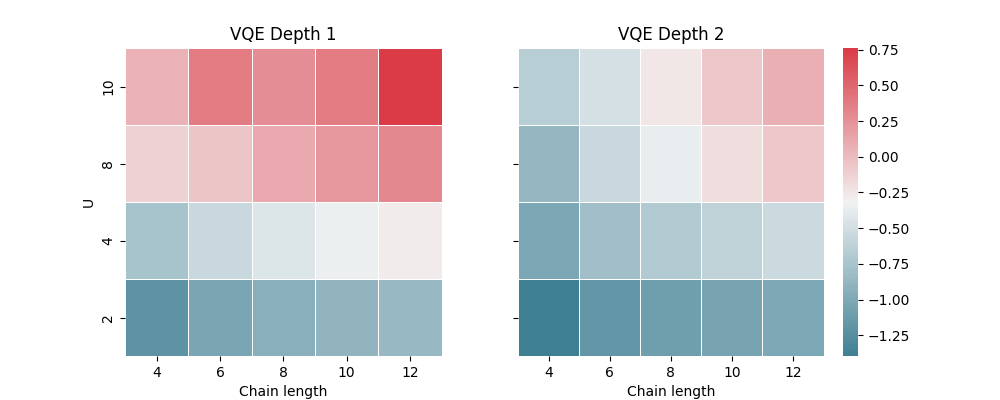}
  \caption{The $\log_{10}$ Frobenius norm error between the VQE XC potential
    functionals and the exact functional for 1D Fermi-Hubbard models at different
    values of interaction strength $U/t$ (where we take $t=1$) and system size. }
\label{fig:1d_XC_fig_c}
\end{figure}

In \cref{fig:1d_XC_fig_c} we provide a
quantitative illustration of the error in the XC functional across varying
interaction strengths and system sizes. To isolate errors from the splining
procedure, we present the Frobenius norm error (on a $\log_{10}$ scale) between
the VQE XC potential functionals and the exact functional with respect to the
discrete values that are computed.
We confirm that VQE functionals decrease in accuracy for bigger systems sizes
with larger $U/t$ parameters and that increasing the number of VQE layers can
significantly improve the quality of the underlying functional, especially in
the high $U/t$ ($\approx 10$) and large system size ($\approx 1 \times 12$)
parameter regime of the 1D Hubbard model. 

When $U/t$ is smaller, i.e $\leq 4$,
it is sufficient to use a VQE functional of depth 1 for all system sizes up to
$1 \times 12$, as the overall error metric is on a similar scale as that for the
VQE functionals of depth 2. Indeed, our results suggest that the XC functional
in 1D is more sensitive to increases in the interaction strength than similar
increases in the system size, indicating that there exists legitimate regions of
applicability for low-depth VQE functionals in DFT computations. This bias is
due to the structure of the ansatz circuit for VQE, that uses the non-interacting groundstate at $U=0$ as a starting state. Potential improvements could be made by varying this ansatz.


\begin{figure}[t!] \centering
  \includegraphics[width=\linewidth]{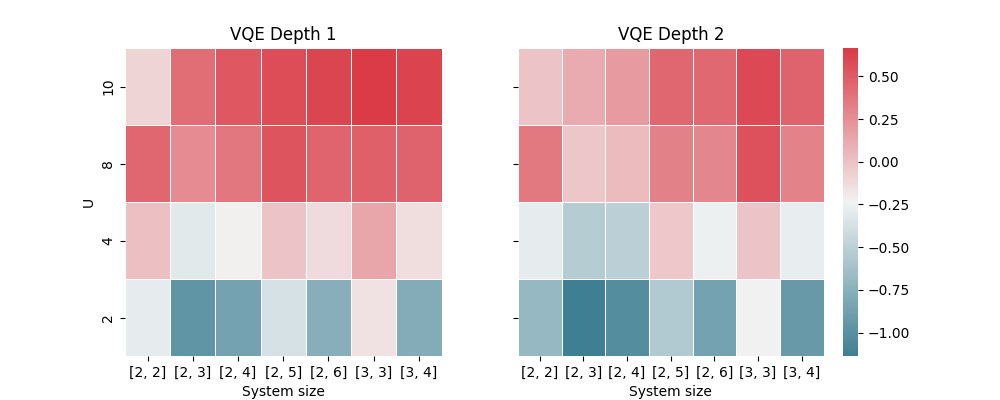}
  \caption{The $\log_{10}$ Frobenius norm error between the VQE XC potential
    functionals and the exact functional for 2D Fermi-Hubbard models at different
    values of interaction strength $U/t$ and system size $[n_x, n_y]$.}
  \label{fig:2d_fig_aII}
\end{figure}

In \cref{fig:2d_fig_aII} we examine the effect that VQE depth has across
a range of 2D system sizes. Similar to 1D, it is clear that the leading indicator for quality in
the VQE functional is the variational depth at which the quantum simulation is
performed. While we observe that the 2D results do not see the same quality of
improvements at VQE depth 2 as the 1D systems show, most notably in the $U \geq 8$
regions, we nevertheless do see consistent improvements over depth 1. This
suggests that 2D VQE functionals may benefit, especially at larger interaction
strengths, from using slightly larger circuit depths, or varying the circuit
ansatz as discussed above. Similar to the 1D case, we see that for $U$ values
in the region of $U \leq 4$ it is sufficient to use VQE functionals of depth 1
in favour of depth 2 as they retain a similar accuracy.

\section{QEDFT in 1D}\label{app:qedft_1d}

\subsection{Emulation}

\begin{figure}[t!] \centering
  \includegraphics[scale=0.6]{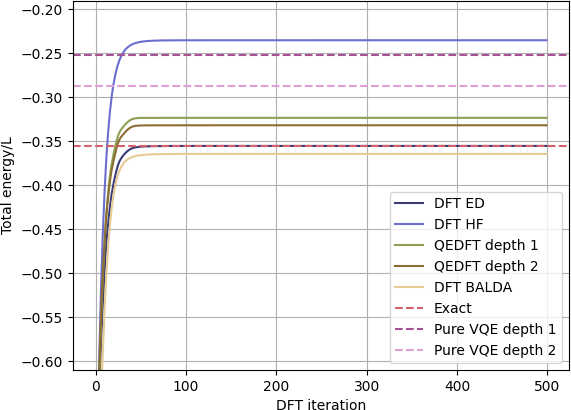}
  \includegraphics[scale=0.6]{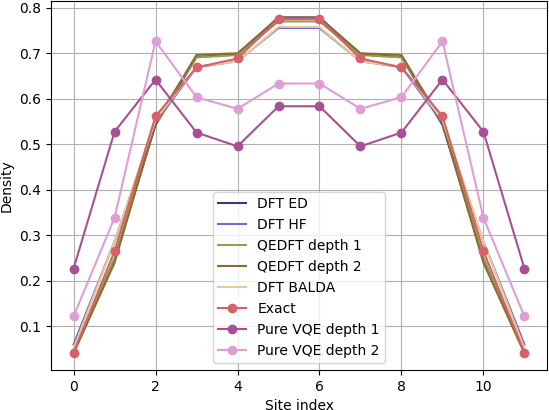}
  \caption{The $1 \times 12$ Fermi Hubbard model for $U=4$ and at
    quarter-filling with an external quadratic potential. The top
    plot illustrates the convergence of the total energy per lattice
    site within the DFT loop. The bottom plot displays the
    density as predicted using different DFT functionals and the
    corresponding exact methods. }
\label{fig:1d_fig_a}
\end{figure}

In \cref{fig:1d_fig_a}  we present QEDFT, DFT, VQE, and exact results for the $1 \times 12$
Fermi-Hubbard model with an external quadratic potential $v_{i}^{ext}= (i-1)^2/L$ with $L = 12$.
The model parameters are $U = 4$, $t =1$,
and $N_e = 6$ (i.e quarter-filling). The exact results are found
using exact diagonalisation of the full system.


\cref{fig:1d_fig_a} shows the convergence of the energy and the
corresponding final density for all methods, with the final errors displayed in~\cref{tab:1d_a_error_metrics}. Considering the energetics, the
QEDFT method clearly achieves better accuracy than both DFT HF alone
(Hartree Fock - which by definition has no discontinuity) and pure quantum VQE alone
and converges after $\sim 100$ iterations to $\delta$, the DFT
self-consistency criteria, within machine precision. \cref{fig:1d_fig_a} also highlights that despite only
having access to qualitatively accurate XC energy functionals via VQE
simulations it is possible to outperform popular classical approximations and/or
purely quantum approaches. 

Similarly the density also converges to machine precision in the DFT
self-consistency criteria. The pure VQE methods struggle to reproduce the sharp central peak at the
centre of the chain generated by the external quadratic potential and is also
considerably erroneous near the boundaries. The QEDFT approach captures these
features well over the entire chain and are qualitatively indistinguishable from
the state-of-the-art classical methods. Moreover in
\cref{tab:1d_a_error_metrics} the errors of the groundstate
density from QEDFT are comparable to all of the classical methods. Notably the
QEDFT density at depth 1 is more accurate than at depth 2 and we attribute this
difference to the interpolation scheme used for the XC energy and potential,
which we expect to disappear at higher depths. From this result we see that
QEDFT not only can predict the groundstate energy accurately, it is also well
suited to predicting the groundstate density, and in particular it is far
superior than using pure VQE alone.

\begin{table}[!tb]
  \centering
  \begin{tabular}{| l | l | l | }
    \hline

    Method & $\Delta n$ & $\Delta E$ \\ \hline
    Pure VQE depth 1 & $0.635$ & $0.10$ \\ \hline
    Pure VQE depth 2 & $0.388$ & $0.067$ \\ \hline
    QEDFT depth 1 & $0.043$ & $0.031$ \\ \hline
    QEDFT depth 2 & $0.06$ & $0.023$ \\ \hline
    DFT HF & $0.052$ & $0.11$ \\ \hline
    DFT BALDA & $0.044$ & $0.009$ \\ \hline
    DFT ED & $0.050$ & $0.0005$ \\

    \hline
  \end{tabular}
  \caption{The Frobenius error norm in density $\Delta n = \sqrt{\sum_{i}^{L}|n_{i, \text{approx}} - n_{i, \text{EXACT}}|^2}$ and error in energy $\Delta E = |E_{\text{approx}} - E_{\text{EXACT}}|$ after 500 DFT iterations with respect to the exact groundstate. }
  \label{tab:1d_a_error_metrics}

\end{table}



\begin{figure}[t!] \centering
  \includegraphics[width=\linewidth]{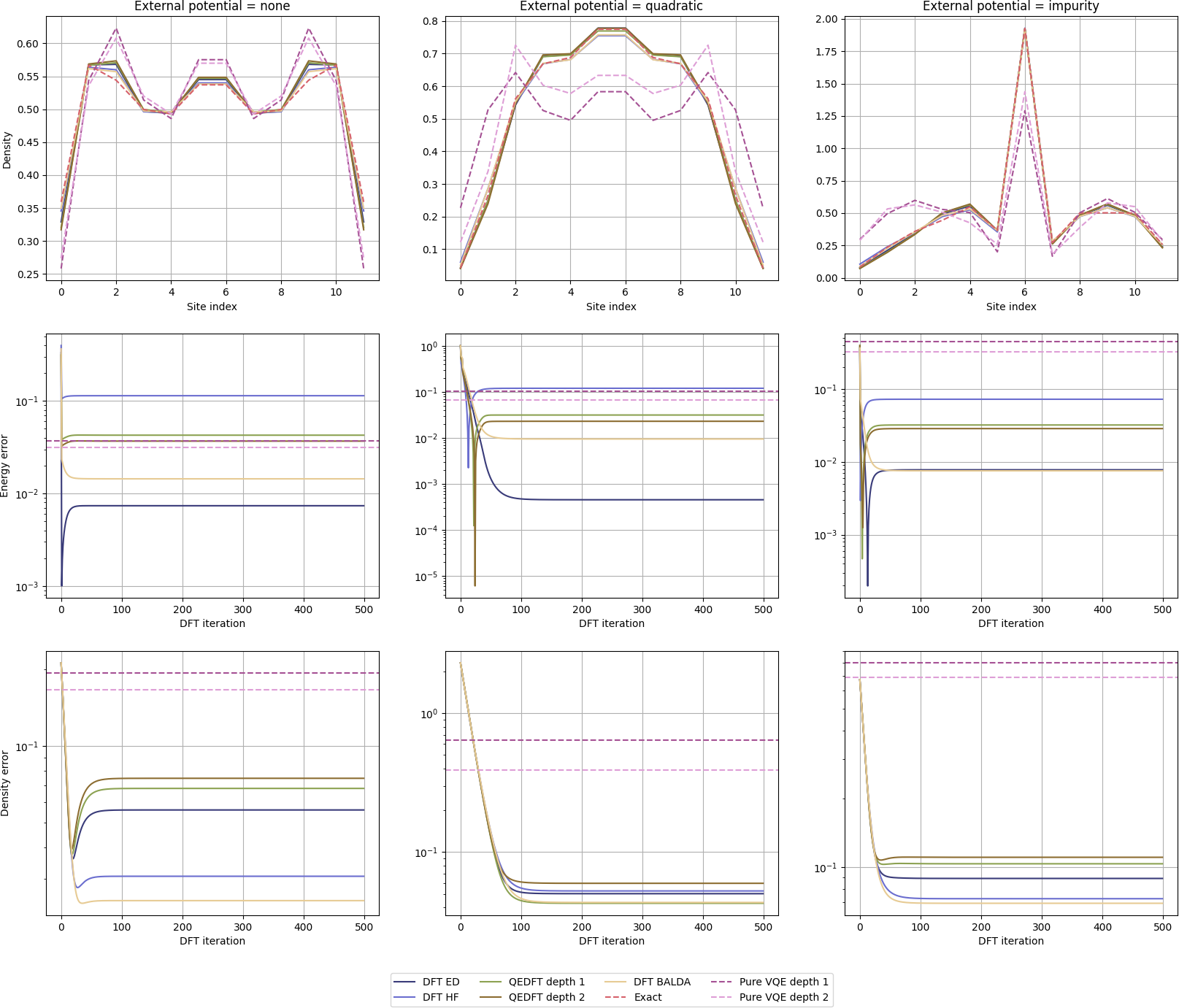}
  \caption{The $1 \times 12$ Fermi Hubbard model for $U=4$ and at
    quarter-filling for different external potentials
    (column-wise). The first row compares the DFT densities compared
    to the exact predictions. The second row monitors the convergence
    of the difference in total energy against the exact solution over
    the duration of the DFT computation. The third row monitors the
    convergence of the difference in total density against the exact
    solution over the duration of the DFT computation.}
\label{fig:1d_fig_b}
\end{figure}

We extend our analysis to understand how QEDFT deals with the impact of changing
the type of external potential for the $1 \times 12$ Fermi-Hubbard model. In
\cref{fig:1d_fig_b} we highlight the DFT groundstate properties and how they
converge for \emph{(i)} no external potential, \emph{(ii)} a confining quadratic
potential and \emph{(iii)} a single impurity potential placed near the centre of
the chain. In the top panel of \cref{fig:1d_fig_b} we show the groundstate
density prediction across all potentials for all methods. The observations we
made for the quadratic potential also apply to the case of no potential as well
as an impurity in the centre of the system, i.e the densities from QEDFT are qualitatively
always more accurate than pure quantum VQE and comparable to the classical
state-of-the-art. 

Focusing on the case of no potential, which is equivalent to the
homogeneous problem from which the XC functionals are generated, we emphasise
that even here the QEDFT approach yields better groundstate density estimates
than VQE alone. That is to say, by first doing a VQE computation
and then doing a QEDFT computation using the VQE data we are able to arrive at a
more accurate description of the groundstate density. This suggest
a wide range of applicability of the QEDFT approach for general
homogeneous and inhomogeneous 1D Fermi-Hubbard Hamiltonians, where the QEDFT
functionals can provide an avenue for using small quantum machines to improve
over pure VQE as well as classical approximations.

\begin{figure}[t!] \centering
  \includegraphics[width=\linewidth]{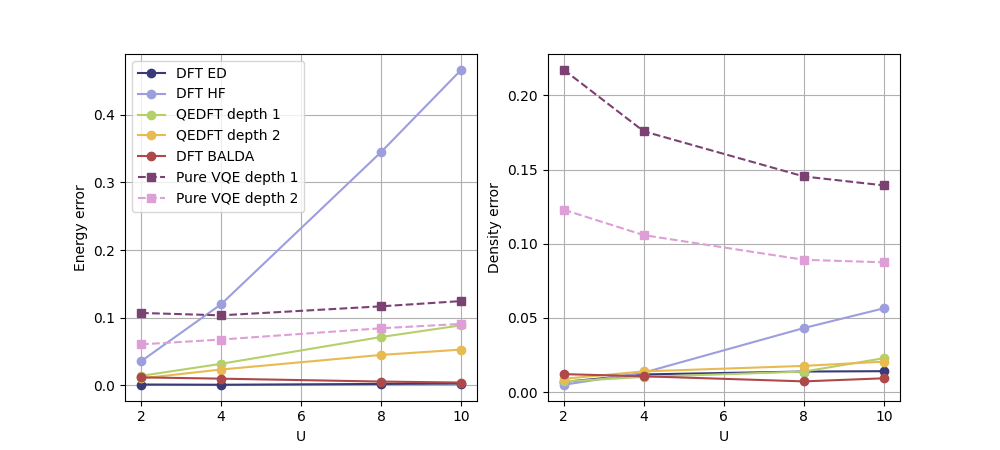}
  \caption{The $1 \times 12$ Fermi Hubbard model at quarter-filling
    with a quadratic external potential and its respective errors in
    energy (left panel) and density (right panel) for different
    interaction strengths. All simulations use DFT apart from the
    dashed lines.}
\label{fig:1d_fig_c}
\end{figure}

The strength of interactions $U/t$ determines how correlated the groundstate of the Fermi-Hubbard
model is, which is largely captured within the XC potential $V_{XC}$. As $U/t$
is increased, classical HF theory breaks down in its ability to describe
groundstate properties as it explicitly sets the XC potential to zero and
reduces the Fermi-Hubbard problem into a purely single particle problem in the
absence of exchange and correlation. In \cref{fig:1d_fig_c} we examine the
effect of increasing $U/t$ on the groundstate energy and density for the
$1\times 12$ system at quarter-filling within the same confining quadratic
potential. This is a direct probe of how well the QEDFT functional can
encapsulate XC effects.

In \cref{fig:1d_fig_a} and \cref{fig:1d_fig_b}, we noticed that while QEDFT
always outperforms VQE in accuracy for both energy and density metrics,
it competed with all classical methods, importantly including HF for the
density. In \cref{fig:1d_fig_c} we see that when $U/t \geq 4$, QEDFT consistently
is more accurate than HF for both the energy and the density. Indeed, as $U/t$
increases, the HF approximation becomes less well justified due to the presence
of correlations. On
the other hand, when $U/t$ is small the presence of interactions is weaker and
the validity of HF is evidenced by its capacity to predict the groundstate
density, despite the energy being vastly overestimated. Importantly, this acts
as a demonstration that QEDFT functionals, even those which are generated from
low-depth VQE circuits, are able to reproduce essential features of correlation
that are needed to describe the Mott insulating phase of the Fermi-Hubbard
model. 

\begin{figure}[t!] \centering
  \includegraphics[width=0.85\linewidth]{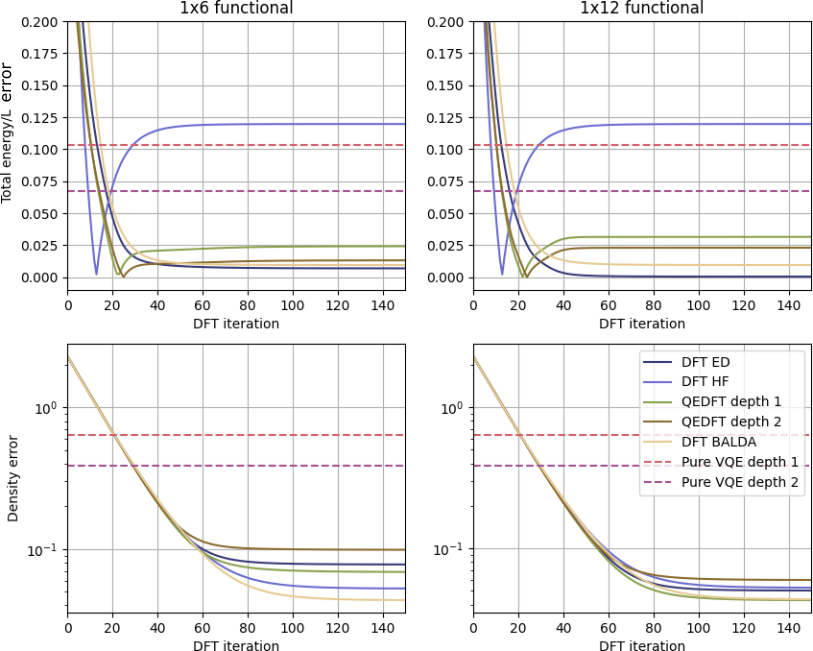}
  \caption{Performance of the finite size scaling of XC functionals in the convergence of the
    total energy error (top panel) and density error (bottom panel) with respect
    to the exact solution for the quadratic external potential. Left column are the
    results produced with $1 \times 6 $ XC potential. Right column are the results
    produced with $1 \times 12$ XC potential.}
\label{fig:1d_fig_d}
\end{figure}




\subsubsection{Finite size scaling}
Another pertinent question related to the scalability of the QEDFT method is if
the performance of the functional improves as a function of system size. We
probe this question by using functionals generated on $1\times6$ and $1\times12$
homogeneous systems applied to $1\times12$ inhomogeneous models with a quadratic
potential. In the top panel of \cref{fig:1d_fig_d} we see that the error in the
energy when using different functionals manifestly improves for the ED
functional, indicating that the quality of the solution gets better by using DFT
functionals which are generated on larger grids. We note that the DFT HF and DFT
BALDA solutions in both of these cases are the same, as they are not sensitive
to system size. Notably, the QEDFT energetics do not demonstrably improve,
rather slightly worsen, when increasing the size of the system used to obtain the functional. We attribute this
behaviour to the worsening of the VQE algorithm with system size at fixed depth,
and expect this to be resolved by increasing the ansatz depth. 

However, in the bottom panel
of \cref{fig:1d_fig_d} we find that the total error of the density undeniably
improves as the system on which the functional is used increases. There is a
systematic reduction of the error for each approximation, with BALDA being the
most accurate method to predict the density. We note that the quality of these
results will be dependent on the inhomogeneous models studied and their
parameter settings, but largely we have identified that the above trends hold
across the various systems we have examined.

\subsection{Hardware}

\begin{figure}[t!] \centering
  \includegraphics[width=\linewidth]{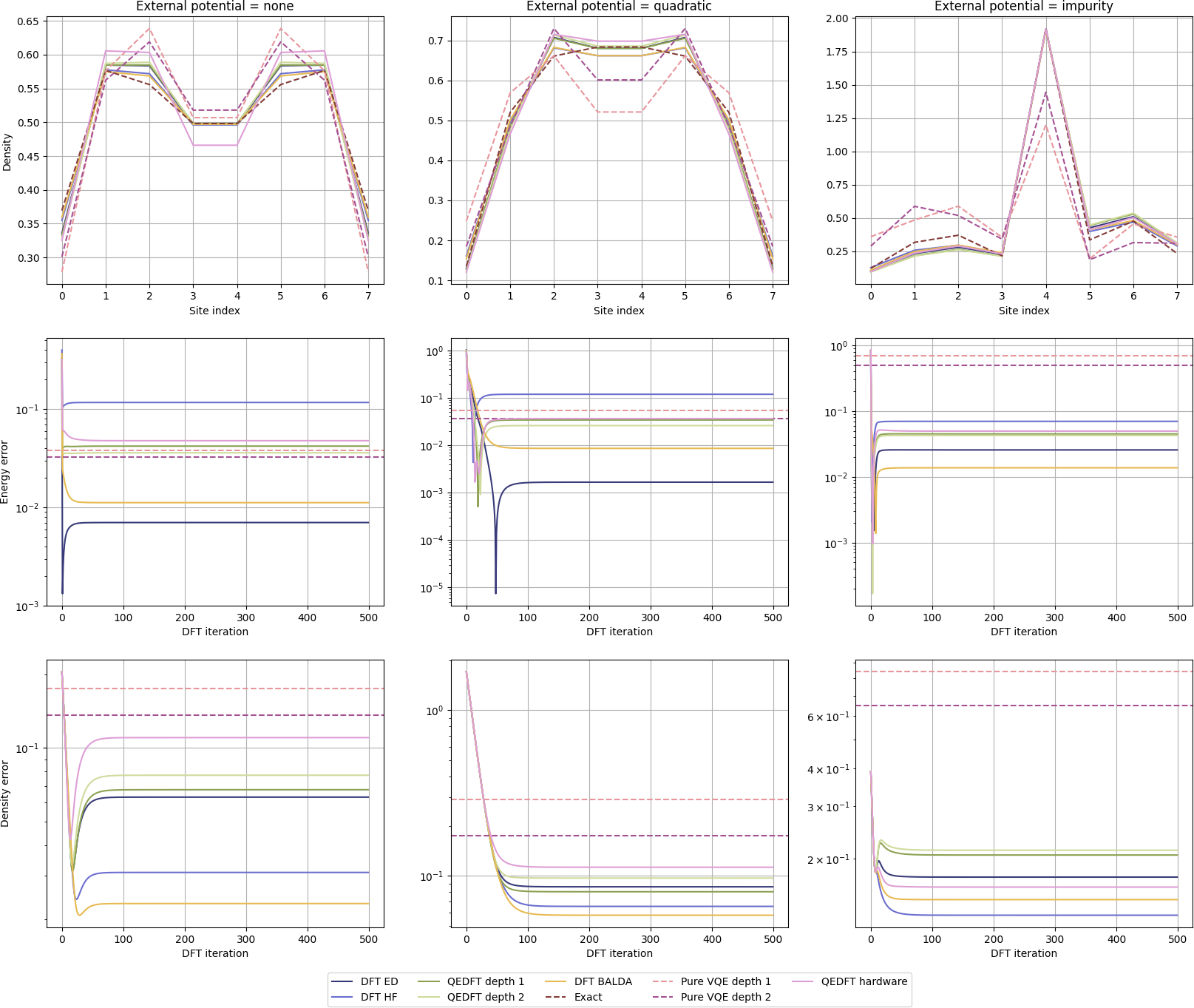}
  \caption{The $1 \times 8$ Fermi Hubbard model for $U=4$ and at quarter- filling
    for different external potentials (column-wise) including results on hardware. The first row
    compares the DFT densities compared to the exact predictions. The second row
    monitors the convergence of the difference in total energy against the exact
    solution over the duration of the DFT computation. The third row monitors
    the convergence of the difference in total density against the exact
    solution over the duration of the DFT computation.}
\label{fig:1d_HW_fig_b}
\end{figure}

We analyse the effect of this noise by implementing first the QEDFT functional
for the $1\times8$ model for different homogeneous and inhomogeneous $1\times8$
Fermi-Hubbard models. We choose this model as it represents the case where the
XC potential possesses significant hardware noise, in order to assess how this
manifests in the DFT results. The results are shown in \cref{fig:1d_HW_fig_b}.
Despite the presence of noise, the hardware data achieves more accurate results
than brute force simulated VQE at depth 1 and 2 in all scenarios, for both the
energy and density. The QEDFT hardware energetics consistently achieve accuracies
relative to the true energy between $10^{-1}$ and $10^{-2}$, which is
significantly better than the classical HF results. While the QEDFT hardware densities
do not outperform the classical HF densities, they are certainly similar in
accuracy to the QEDFT simulated functionals, which can be improved by including
more variational layers. Indeed, even though there is the clear presence of
noise in the QEDFT hardware functional, manifest in both the raw data as well its
interpolation, we show that its subsequent application in a DFT
simulation is largely oblivious to these details for the model parameters we have considered.

\section{QEDFT in 2D}\label{app:qedft_2d}

\subsection{Emulation}

\begin{figure}[t!] \centering
  \includegraphics[width=\linewidth]{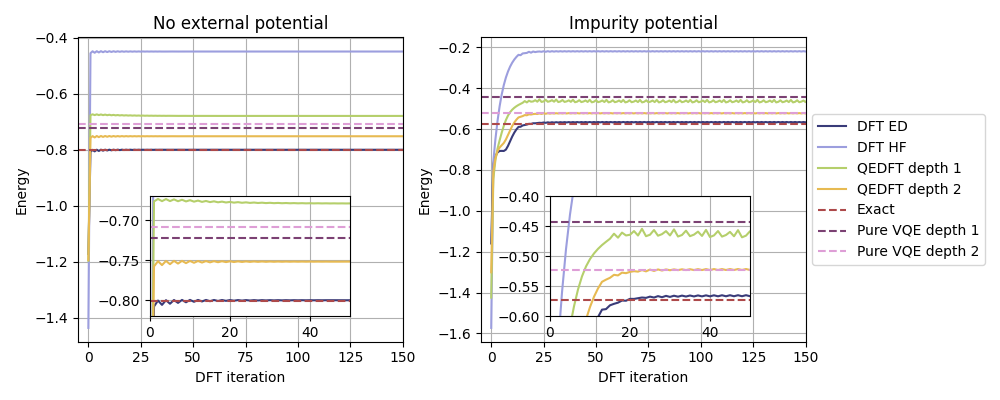}
  \caption{The DFT energy convergence $3 \times 3$ Fermi Hubbard model for $U/t=10$
    and at quarter-filling for no external potential and an impurity
    potential at the $(1,2)$ site. Dashed lines represent non DFT calculations,
    which are either the exact solution or pure VQE emulations.}
\label{fig:2d_fig_b_i}
\end{figure}

We now proceed with the DFT analysis of Fermi-Hubbard models in 2D,
beginning with the energy convergence of the DFT energy, presented in
\cref{fig:2d_fig_b_i} for the $3 \times 3$ system with $U/t = 10$ at quarter
filling. We consider the homogeneous model as well as the inhomogeneous system
with an impurity potential embedded in a random background at the $(1, 2)$ site. 

For the
homogeneous model we observe that the convergence of DFT energy is initially
unstable, as evidenced by the jagged spikes for the first 50 iterations. This
behaviour is present in all functionals studied, indicating that this
instability is not associated with the particular form of the XC potential but
rather the stability of the single particle Kohn-Sham approach for the chosen
parameter regime. A high density mixing parameter ($\alpha = 0.95$) ensures that
early instabilities cannot propagate towards erroneous fixed points. However, inhomogeneities introduce more noticeable instabilities in
the DFT algorithm, as evidenced by the jaggedness of the energy convergence
around a fixed value. While not severe in this case, a combination of more
accurate XC interpolation methods and higher density mixing parameters should
remove convergence to multiple fixed points.

It is clear that in all cases using a VQE approach is
superior to Hartree Fock, which for example has an error in the converged energy of
$\approx 44 \%$ for the homogeneous model, in contrast to the most accurate VQE method (QEDFT at depth 2)
$\approx 6\%$. Indeed, we further see that QEDFT at depth 2 attains an
accuracy closer to the exact solution than pure VQE alone, which is not achieved
at QEDFT depth 1. In contrast to the 1D results, this suggests that going beyond depth 1
QEDFT functionals is not only beneficial, but also necessary, to improve upon
using pure VQE. Finally, the ED functional produces the groundstate
energy with an error of $\approx 0.1\%$, indicating that the local density
approximation when sampled at the converged density produces a good, but not exact, estimate of
the groundstate energy for 2D systems also. 

\begin{figure}[t!] \centering
  \includegraphics[width=\textwidth]{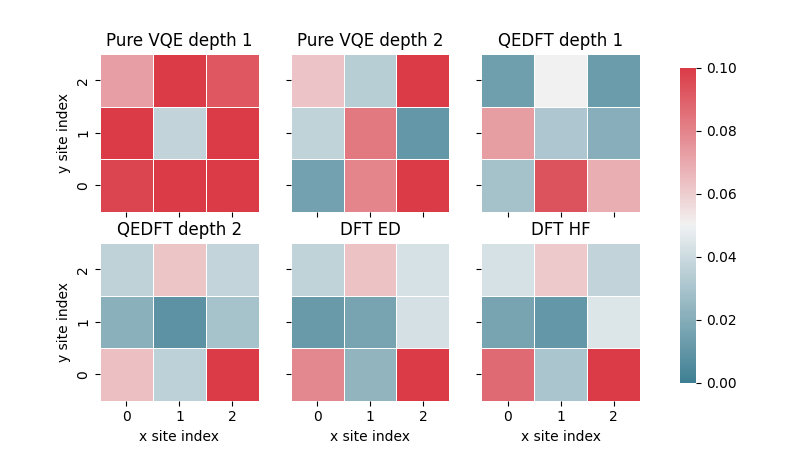}
  \caption{The per site density percentage error \cref{eq:dens_percent} for the
    $3 \times 3$ Fermi Hubbard model with impurity at site $(1,2)$ for $U/t=8$. Pure VQE methods are highlighted in the
    top panel, while the other methods use DFT.}
\label{fig:2d_fig_b_ii}
\end{figure}

In \cref{fig:2d_fig_b_ii} we analyse converged groundstate densities for the
inhomogeneous $3\times3$ system by presenting the site density percentage error with the
groundstate density,
\begin{equation}
  \Delta n_{ij} = \frac{|n_{ij}^{\text{approx}} - n_{ij}^{\text{exact}}|}{n_{ij}^{\text{exact}}}.
  \label{eq:dens_percent}
\end{equation}

We arrive at a set of slightly different conclusions for the DFT groundstate
densities as compared to the energies. For the densities, the DFT methods are
always more reliable than pure VQE, which produces errors per site that
are mostly all above $0.1 \%$. QEDFT at depth 2 highly resembles the
DFT ED and DFT HF solutions, indicating that low-depth QEDFT methods can compete
with classical state-of-the-art approaches for predicting the groundstate
density in 2D. Moreover, despite potential errors arising in the interpolation
of the XC potential due to rogue data points, i.e as seen in \cref{fig:fig5}
for the $3\times3$ case, we see that the DFT algorithm can be robust against
these, depending on where the density is sampled. If the curvature of the
underlying function is non-trivial, which it is for the $3\times3$ system,
improvements can be made by using denser meshes to assist the interpolated XC
potential and prevent DFT from driving the density to a result that is further away
from the exact one, but are not necessary for this inhomogeneous model. 

\begin{figure}[t!] \centering
  \includegraphics[width=\linewidth]{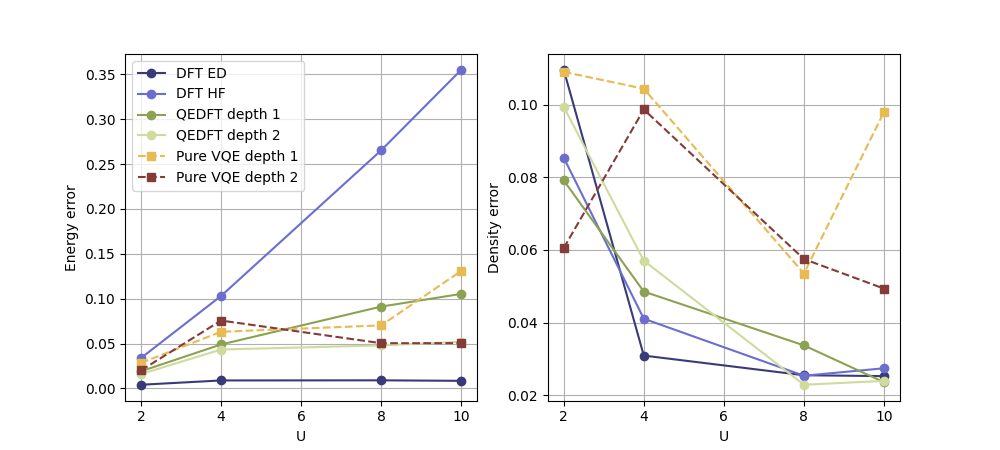}
  \caption{The $3 \times 3$ Fermi Hubbard model at quarter-filling with an
    impurity potential at the $(1, 2)$ site and its respective errors in energy
    (left panel) and density (right panel) for different interaction strengths.
    All simulations use DFT apart from the dashed lines.}
\label{fig:2d_fig_c}
\end{figure}

To quantify this error further we study the effect of varying $U/t$ on the
absolute errors with respect to the groundstate energies and densities for the
quarter-filled $3\times3$ inhomogeneous impurity system, illustrated in
\cref{fig:2d_fig_c}. The results for the energetics are reminiscent of those
presented for 1D systems in \cref{fig:1d_fig_c}. The density results,
however, deviate significantly from what we observed in 1D. Indeed, as $U/t$
increases the density error decreases, which is opposite to the trend we observed
for the 1D quadratic potential. Even at low $U$, the DFT ED functional does not
reproduce the exact solution, indicating that the LDA approximation for
the 2D Fermi Hubbard model at low $U/t$ is not as reliable as in 1D for
predicting the groundstate density.
Generally speaking, the DFT methods, and in particular the QEDFT functionals,
always provide more reliable densities than pure VQE alone across all values of
$U/t$, except at $U/t=2$, where pure VQE at depth 2 is the most accurate
approach.



\subsubsection{Finite size scaling}
\begin{figure}[t!] \centering
  \includegraphics[width=\linewidth]{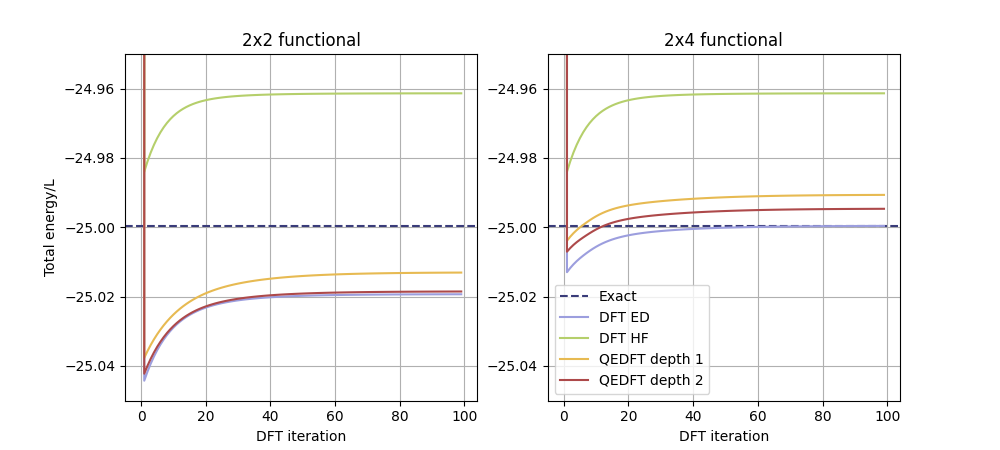}
  \caption{Comparing the performance of the $2\times 2$ and $2\times 4$ XC functional, for a $2\times 4$ model with an impurity at site $(1, 2)$. The energy per site is plotted against the DFT iteration.}
\label{fig:2d_extrap}
\end{figure}

We now revisit examining in 2D how well QEDFT scales as the functionals used are
increased in the system size. In
\cref{fig:2d_extrap} we show that studying the inhomogeneous $2\times4$
Fermi-Hubbard system with an impurity at site $(1,2)$ improves considerably when
going from a functional generated on a $2\times2$ lattice, as compared to using
a functional from a $2\times4$ system. Notably, the DFT ED results converge to
the exact method using the $2\times4$ functional, demonstrating the utility of
how the functional can scale with system size. Moreover, the QEDFT energetics also
converge to values which are closer to the exact value, but are not as
close as DFT ED. The pure VQE energetics are not shown, as they incur an error
of approximately $\sim 20\%$, as opposed to the DFT methods, which are all
within the region of less than $\sim 0.2 \%$. In contrast to the 1D results in~\cref{fig:1d_fig_d}, for 2D we did not observe a
significant improvement related to the density in 2D.

\subsection{Hardware}



\begin{figure}[t!] \centering
  \includegraphics[width=\linewidth]{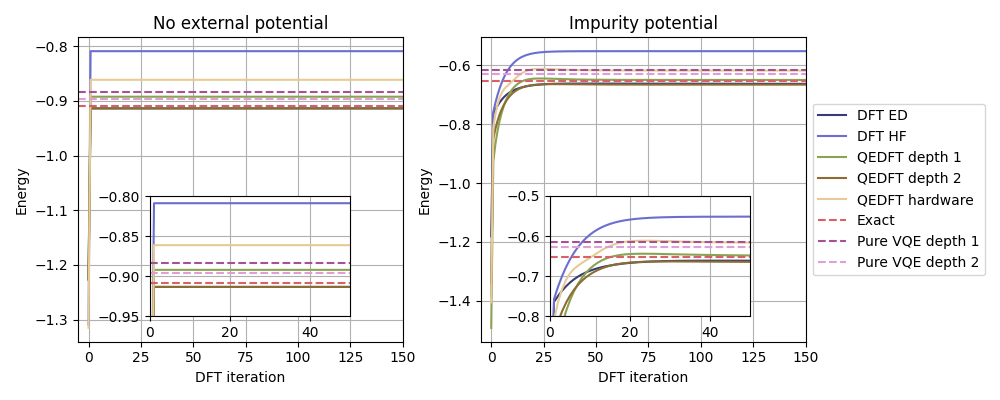}
  \caption{The DFT energy convergence $2 \times 4$ Fermi Hubbard model for
    $U/t=4$ and at quarter-filling for no external potential and an impurity
    potential at the $(1,1)$ site using the hardware data. Dashed lines
    represent non DFT calculations, which are either the exact solution or pure
    VQE emulations.}
\label{fig:2d_HW_fig_a}
\end{figure}

\begin{figure}[t!] \centering
  \includegraphics[width=\linewidth]{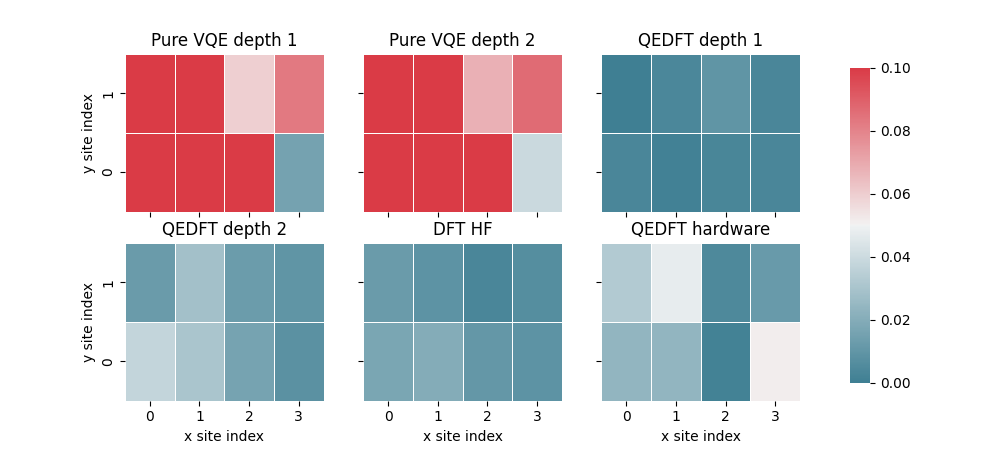}
  \caption{The per site density percentage error
    \cref{eq:dens_percent} for the quarter-filled $2 \times 4$ Fermi
    Hubbard model with impurity at site $(1,2)$ for $U/t=4$ including
    hardware data.}
\label{fig:2d_HW_fig_b}
\end{figure}

We study the $2\times4$ Fermi-Hubbard system using the hardware functional
without an external potential and in the presence of a single impurity at the
$(1,1)$
site embedded in a disordered external potential. The results are shown in
\cref{fig:2d_HW_fig_a} and \cref{fig:2d_HW_fig_b}.
For both the
homogeneous and inhomogeneous models we see that that all DFT methods are stable, in contrast to the
$3\times3$ simulations. The origin of these differences is possibly related to
the behaviour of the XC functional near half-filling as we presented in
\cref{fig:fig5}, which is noticeably worse for the square $3\times3$
lattice than for rectangular lattices. Indeed, this indicates that errors
originating from the splining of the XC functional may have less of an impact on the stability of the DFT
algorithm than absolute error of the raw data to the exact reference values.

We also find that the DFT hardware energetics, while consistently better than classical HF, is either slightly worse or essentially no better
than exact emulations of pure VQE. As there is no hardware data for the direct
implementation of VQE for these target problems we cannot directly compare DFT
VQE to pure VQE on hardware, but anticipate DFT VQE to outperform the latter as
we consistently observed that in emulation.

The hardware data is also used to assess the accuracy of DFT for the
groundstate density of the inhomogeneous $2\times4$ problem in
\cref{fig:2d_HW_fig_b}. Interestingly, the hardware functional produces
groundstate densities a lot closer than pure VQE emulation, having a per-site
percentage error of less than $0.1\%$ for all sites, while most sites predicted
with the VQE emulation have errors greater than $0.1\%$. As was observed
for the emulations of the $3\times3$ model, the DFT
methods based on VQE are comparable to classical HF, which is also true for the
hardware functional in this case. Notably, the QEDFT depth 1 densities produce
results that are closer to the true solution than QEDFT depth 2. Indeed, this
is also reflected in the accuracy of the energetics, and is related to the
quality of the interpolation method performed. Due to the finite data set over
which the XC functional is constructed, it is possible to introduce error at
higher depths within the interpolated region if the spacing is not large
enough, with respect to lower-depth results. Therefore, we anticipate that
classical improvements to the interpolation protocol of QEDFT should make it
possible to obtain groundstate properties closer to the true results.

\printbibliography

@article{baker_density_2020,
	title = {Density functionals and {Kohn}-{Sham} potentials with minimal wavefunction preparations on a quantum computer},
	volume = {2},
	issn = {2643-1564},
	url = {https://link.aps.org/doi/10.1103/PhysRevResearch.2.043238},
	doi = {10.1103/PhysRevResearch.2.043238},
	language = {en},
	number = {4},
	urldate = {2022-04-04},
	journal = {Physical Review Research},
	author = {Baker, Thomas E. and Poulin, David},
	month = nov,
	year = {2020},
	pages = {043238},
}

@article{hatcher_method_2019,
	title = {A {Method} to {Calculate} {Correlation} for {Density} {Functional} {Theory} on a {Quantum} {Processor}},
	url = {http://arxiv.org/abs/1903.05550},
	urldate = {2022-04-04},
	journal = {arXiv:1903.05550 [quant-ph]},
	author = {Hatcher, Ryan and Kittl, Jorge A. and Bowen, Christopher},
	month = mar,
	year = {2019},
}

@article{ma_quantum_2020,
	title = {Quantum simulations of materials on near-term quantum computers},
	volume = {6},
	issn = {2057-3960},
	url = {http://www.nature.com/articles/s41524-020-00353-z},
	doi = {10.1038/s41524-020-00353-z},
	language = {en},
	number = {1},
	urldate = {2022-04-05},
	journal = {npj Computational Materials},
	author = {Ma, He and Govoni, Marco and Galli, Giulia},
	month = dec,
	year = {2020},
	pages = {85},
}

@article{rossmannek2021quantum,
  title={Quantum HF/DFT-embedding algorithms for electronic structure calculations: Scaling up to complex molecular systems},
  author={Rossmannek, Max and Barkoutsos, Panagiotis Kl and Ollitrault, Pauline J and Tavernelli, Ivano},
  journal={The Journal of Chemical Physics},
  volume={154},
  number={11},
  pages={114105},
  year={2021},
  publisher={AIP Publishing LLC},
  url = {http://dx.doi.org/10.1063/5.0029536},
  DOI = {10.1063/5.0029536},
}

@article{Desh_qdft,
  title = {Levy-Lieb embedding of density-functional theory and its quantum kernel: Illustration for the Hubbard dimer using near-term quantum algorithms},
  author = {Pemmaraju, C. D. and Deshmukh, Amol},
  journal = {Phys. Rev. A},
  volume = {106},
  issue = {4},
  pages = {042807},
  numpages = {10},
  year = {2022},
  publisher = {American Physical Society},
  doi = {10.1103/PhysRevA.106.042807},
  url = {https://link.aps.org/doi/10.1103/PhysRevA.106.042807}
}

@article{senjean2022toward,
	title={{Toward density functional theory on quantum computers?}},
	author={Bruno Senjean and Saad Yalouz and Matthieu Saubanère},
	journal={SciPost Phys.},
	volume={14},
	pages={055},
	year={2023},
	publisher={SciPost},
	doi={10.21468/SciPostPhys.14.3.055},
	url={https://scipost.org/10.21468/SciPostPhys.14.3.055},
}

@article{montanaro2020compressed,
  title={Compressed variational quantum eigensolver for the Fermi-Hubbard model},
  author={Montanaro, Ashley and Stanisic, Stasja},
  journal={arXiv:2006.01179 [quant-ph]},
  year={2020},
  doi = {10.48550/ARXIV.2006.01179},
  url = {https://arxiv.org/abs/2006.01179},
}

@article{sanvitoML,
  title = {Machine learning density functional theory for the Hubbard model},
  author = {Nelson, James and Tiwari, Rajarshi and Sanvito, Stefano},
  journal = {Phys. Rev. B},
  volume = {99},
  issue = {7},
  pages = {075132},
  numpages = {5},
  year = {2019},
  publisher = {American Physical Society},
  doi = {10.1103/PhysRevB.99.075132},
  url = {https://link.aps.org/doi/10.1103/PhysRevB.99.075132}
}

@article{DM21,
author = {James Kirkpatrick  and Brendan McMorrow  and David H. P. Turban  and Alexander L. Gaunt  and James S. Spencer  and Alexander G. D. G. Matthews  and Annette Obika  and Louis Thiry  and Meire Fortunato  and David Pfau  and Lara Román Castellanos  and Stig Petersen  and Alexander W. R. Nelson  and Pushmeet Kohli  and Paula Mori-Sánchez  and Demis Hassabis  and Aron J. Cohen },
title = {Pushing the frontiers of density functionals by solving the fractional electron problem},
journal = {Science},
volume = {374},
number = {6573},
pages = {1385-1389},
year = {2021},
doi = {10.1126/science.abj6511},
URL = {https://www.science.org/doi/abs/10.1126/science.abj6511}}

@article{burkeML,
	author = {Bogojeski, Mihail and Vogt-Maranto, Leslie and Tuckerman, Mark E. and M{\"u}ller, Klaus-Robert and Burke, Kieron},
	da = {2020/10/16},
	doi = {10.1038/s41467-020-19093-1},
	id = {Bogojeski2020},
	isbn = {2041-1723},
	journal = {Nature Communications},
	number = {1},
	pages = {5223},
	title = {Quantum chemical accuracy from density functional approximations via machine learning},
	ty = {JOUR},
	url = {https://doi.org/10.1038/s41467-020-19093-1},
	volume = {11},
	year = {2020}}

@article{BurkeMLII,
  title = {Kohn-Sham calculations with the exact functional},
  author = {Wagner, Lucas O. and Baker, Thomas E. and Stoudenmire, E. M. and Burke, Kieron and White, Steven R.},
  journal = {Phys. Rev. B},
  volume = {90},
  issue = {4},
  pages = {045109},
  numpages = {15},
  year = {2014},
  publisher = {American Physical Society},
  doi = {10.1103/PhysRevB.90.045109},
  url = {https://link.aps.org/doi/10.1103/PhysRevB.90.045109}
}

@article{LDFT_graphene,
  title = {Lattice density-functional theory on graphene},
  author = {Ij\"as, Mari and Harju, Ari},
  journal = {Phys. Rev. B},
  volume = {82},
  issue = {23},
  pages = {235111},
  numpages = {8},
  year = {2010},
  publisher = {American Physical Society},
  doi = {10.1103/PhysRevB.82.235111},
  url = {https://link.aps.org/doi/10.1103/PhysRevB.82.235111}
}

@article{LDFT_BALDA,
  title = {Density Functionals Not Based on the Electron Gas: Local-Density Approximation for a Luttinger Liquid},
  author = {Lima, N. A. and Silva, M. F. and Oliveira, L. N. and Capelle, K.},
  journal = {Phys. Rev. Lett.},
  volume = {90},
  issue = {14},
  pages = {146402},
  numpages = {4},
  year = {2003},
  publisher = {American Physical Society},
  doi = {10.1103/PhysRevLett.90.146402},
  url = {https://link.aps.org/doi/10.1103/PhysRevLett.90.146402}
}

@article{LDFT_original,
  title = {Density-Functional Treatment of an Exactly Solvable Semiconductor Model},
  author = {Gunnarsson, O. and Sch\"onhammer, K.},
  journal = {Phys. Rev. Lett.},
  volume = {56},
  issue = {18},
  pages = {1968--1971},
  numpages = {0},
  year = {1986},
  publisher = {American Physical Society},
  doi = {10.1103/PhysRevLett.56.1968},
  url = {https://link.aps.org/doi/10.1103/PhysRevLett.56.1968}
}

@article{LDFTII,
  title={Testing density-functional approximations on a lattice and the applicability of the related Hohenberg-Kohn-like theorem},
  author={Fran{\c{c}}a, Vivian V and Coe, Jeremy P and D’Amico, Irene},
  journal={Scientific Reports},
  volume={8},
  number={1},
  pages={664},
  year={2018},
  publisher={Nature Publishing Group UK London},
  url = {http://dx.doi.org/10.1038/s41598-017-19018-x},
  DOI = {10.1038/s41598-017-19018-x},
}

@article{carrascal2015hubbard,
  title = {The Hubbard dimer: a density functional case study of a many-body problem},
  volume = {27},
  ISSN = {1361-648X},
  url = {http://dx.doi.org/10.1088/0953-8984/27/39/393001},
  DOI = {10.1088/0953-8984/27/39/393001},
  number = {39},
  journal = {Journal of Physics: Condensed Matter},
  publisher = {IOP Publishing},
  author = {Carrascal,  D J and Ferrer,  J and Smith,  J C and Burke,  K},
  year = {2015},
  pages = {393001}
}

@article{CAPELLE201391,
title = {Density functionals and model Hamiltonians: Pillars of many-particle physics},
journal = {Physics Reports},
volume = {528},
number = {3},
pages = {91-159},
year = {2013},
issn = {0370-1573},
doi = {https://doi.org/10.1016/j.physrep.2013.03.002},
url = {https://www.sciencedirect.com/science/article/pii/S0370157313000975},
author = {Klaus Capelle and Vivaldo L. Campo}
}

@article{Libxc,
title = {Libxc: A library of exchange and correlation functionals for density functional theory},
journal = {Computer Physics Communications},
volume = {183},
number = {10},
pages = {2272-2281},
year = {2012},
issn = {0010-4655},
doi = {https://doi.org/10.1016/j.cpc.2012.05.007},
url = {https://www.sciencedirect.com/science/article/pii/S0010465512001750},
author = {Miguel A.L. Marques and Micael J.T. Oliveira and Tobias Burnus}
}

@article{DFT_citations,
author={Haunschild, Robin
and Barth, Andreas
and French, Bernie},
title={A comprehensive analysis of the history of DFT based on the bibliometric method RPYS},
journal={Journal of Cheminformatics},
year={2019},
day={21},
volume={11},
number={1},
pages={72},
issn={1758-2946},
doi={10.1186/s13321-019-0395-y},
url={https://doi.org/10.1186/s13321-019-0395-y}}

@article{DFT_RMP,
  title = {Density functional theory: Its origins, rise to prominence, and future},
  author = {Jones, R. O.},
  journal = {Rev. Mod. Phys.},
  volume = {87},
  issue = {3},
  pages = {897--923},
  numpages = {27},
  year = {2015},
  publisher = {American Physical Society},
  doi = {10.1103/RevModPhys.87.897},
  url = {https://link.aps.org/doi/10.1103/RevModPhys.87.897}
}

@article{DMFTI_RMP,
  title = {Dynamical mean-field theory of strongly correlated fermion systems and the limit of infinite dimensions},
  author = {Georges, Antoine and Kotliar, Gabriel and Krauth, Werner and Rozenberg, Marcelo J.},
  journal = {Rev. Mod. Phys.},
  volume = {68},
  issue = {1},
  pages = {13--125},
  numpages = {0},
  year = {1996},
  publisher = {American Physical Society},
  doi = {10.1103/RevModPhys.68.13},
  url = {https://link.aps.org/doi/10.1103/RevModPhys.68.13}
}

@article{qsgw,
  title = {Quasiparticle self-consistent $GW$ method: A basis for the independent-particle approximation},
  author = {Kotani, Takao and van Schilfgaarde, Mark and Faleev, Sergey V.},
  journal = {Phys. Rev. B},
  volume = {76},
  issue = {16},
  pages = {165106},
  numpages = {24},
  year = {2007},
  publisher = {American Physical Society},
  doi = {10.1103/PhysRevB.76.165106},
  url = {https://link.aps.org/doi/10.1103/PhysRevB.76.165106}
}

@article{galli22,
  title = {Simulating the Electronic Structure of Spin Defects on Quantum Computers},
  author = {Huang, Benchen and Govoni, Marco and Galli, Giulia},
  journal = {PRX Quantum},
  volume = {3},
  issue = {1},
  pages = {010339},
  numpages = {16},
  year = {2022},
  publisher = {American Physical Society},
  doi = {10.1103/PRXQuantum.3.010339},
  url = {https://link.aps.org/doi/10.1103/PRXQuantum.3.010339}
}

@article{galli22_II,
author = {Huang, Benchen and Sheng, Nan and Govoni, Marco and Galli, Giulia},
title = {Quantum Simulations of Fermionic Hamiltonians with Efficient Encoding and Ansatz Schemes},
journal = {Journal of Chemical Theory and Computation},
volume = {19},
number = {5},
pages = {1487-1498},
year = {2023},
doi = {10.1021/acs.jctc.2c01119},
URL = { https://doi.org/10.1021/acs.jctc.2c01119}
}

@article{vqeI,
	author = {Peruzzo, Alberto and McClean, Jarrod and Shadbolt, Peter and Yung, Man-Hong and Zhou, Xiao-Qi and Love, Peter J. and Aspuru-Guzik, Al{\'a}n and O'Brien, Jeremy L.},
	journal = {Nature Communications},
	number = {1},
	pages = {4213},
	title = {A variational eigenvalue solver on a photonic quantum processor},
	volume = {5},
	year = {2014},
        url = {http://dx.doi.org/10.1038/ncomms5213},
}

@article{vqeII,
doi = {10.1088/1367-2630/18/2/023023},
url = {https://dx.doi.org/10.1088/1367-2630/18/2/023023},
year = {2016},
publisher = {IOP Publishing},
volume = {18},
number = {2},
pages = {023023},
author = {Jarrod R McClean and Jonathan Romero and Ryan Babbush and Alán Aspuru-Guzik},
title = {The theory of variational hybrid quantum-classical algorithms},
journal = {New Journal of Physics}
}

@article{QSE,
  title = {Hybrid quantum-classical hierarchy for mitigation of decoherence and determination of excited states},
  author = {McClean, Jarrod R. and Kimchi-Schwartz, Mollie E. and Carter, Jonathan and de Jong, Wibe A.},
  journal = {Phys. Rev. A},
  volume = {95},
  issue = {4},
  pages = {042308},
  numpages = {10},
  year = {2017},
  publisher = {American Physical Society},
  doi = {10.1103/PhysRevA.95.042308},
  url = {https://link.aps.org/doi/10.1103/PhysRevA.95.042308}
}

@article{dft_QI,
	title = {Computational complexity of interacting electrons and fundamental limitations of density functional theory},
	volume = {5},
	issn = {1745-2473, 1745-2481},
	url = {http://www.nature.com/articles/nphys1370},
	doi = {10.1038/nphys1370},
	language = {en},
	number = {10},
	urldate = {2022-04-04},
	journal = {Nature Physics},
	author = {Schuch, Norbert and Verstraete, Frank},
	month = oct,
	year = {2009},
	pages = {732--735},
}

@article{DFTI,
  title = {Inhomogeneous Electron Gas},
  author = {Hohenberg, P. and Kohn, W.},
  journal = {Phys. Rev.},
  volume = {136},
  issue = {3B},
  pages = {B864--B871},
  numpages = {0},
  year = {1964},
  publisher = {American Physical Society},
  doi = {10.1103/PhysRev.136.B864},
  url = {https://link.aps.org/doi/10.1103/PhysRev.136.B864}
}

@article{DFTII,
  title = {Self-Consistent Equations Including Exchange and Correlation Effects},
  author = {Kohn, W. and Sham, L. J.},
  journal = {Phys. Rev.},
  volume = {140},
  issue = {4A},
  pages = {A1133--A1138},
  numpages = {0},
  year = {1965},
  publisher = {American Physical Society},
  doi = {10.1103/PhysRev.140.A1133},
  url = {https://link.aps.org/doi/10.1103/PhysRev.140.A1133}
}

@article{LDA_QMC,
  title = {Ground State of the Electron Gas by a Stochastic Method},
  author = {Ceperley, D. M. and Alder, B. J.},
  journal = {Phys. Rev. Lett.},
  volume = {45},
  issue = {7},
  pages = {566--569},
  numpages = {0},
  year = {1980},
  publisher = {American Physical Society},
  doi = {10.1103/PhysRevLett.45.566},
  url = {https://link.aps.org/doi/10.1103/PhysRevLett.45.566}
}

@article{dft_func,
author = {Narbe Mardirossian and Martin Head-Gordon},
title = {Thirty years of density functional theory in computational chemistry: an overview and extensive assessment of 200 density functionals},
journal = {Molecular Physics},
volume = {115},
number = {19},
pages = {2315-2372},
year  = {2017},
publisher = {Taylor & Francis},
doi = {10.1080/00268976.2017.1333644},
URL = {https://doi.org/10.1080/00268976.2017.1333644},
}

@article{stanisic2022observing,
  title={Observing ground-state properties of the Fermi-Hubbard model using a scalable algorithm on a quantum computer},
  author={Stanisic, Stasja and Bosse, Jan Lukas and Gambetta, Filippo Maria and Santos, Raul A and Mruczkiewicz, Wojciech and O’Brien, Thomas E and Ostby, Eric and Montanaro, Ashley},
  journal={Nature Communications},
  volume={13},
  number={1},
  pages={5743},
  year={2022},
  publisher={Nature Publishing Group UK London},
  url = {http://dx.doi.org/10.1038/s41467-022-33335-4},
  DOI = {10.1038/s41467-022-33335-4},
}

@article{cade2020strategies,
  title = {Strategies for solving the Fermi-Hubbard model on near-term quantum computers},
  author = {Cade, Chris and Mineh, Lana and Montanaro, Ashley and Stanisic, Stasja},
  journal = {Phys. Rev. B},
  volume = {102},
  issue = {23},
  pages = {235122},
  numpages = {25},
  year = {2020},
  publisher = {American Physical Society},
  doi = {10.1103/PhysRevB.102.235122},
  url = {https://link.aps.org/doi/10.1103/PhysRevB.102.235122}
}

@article{HVA,
  title = {Progress towards practical quantum variational algorithms},
  author = {Wecker, Dave and Hastings, Matthew B. and Troyer, Matthias},
  journal = {Phys. Rev. A},
  volume = {92},
  issue = {4},
  pages = {042303},
  numpages = {10},
  year = {2015},
  publisher = {American Physical Society},
  doi = {10.1103/PhysRevA.92.042303},
  url = {https://link.aps.org/doi/10.1103/PhysRevA.92.042303}
}

@article{clinton2022towards,
  title = {Towards near-term quantum simulation of materials},
  volume = {15},
  ISSN = {2041-1723},
  url = {http://dx.doi.org/10.1038/s41467-023-43479-6},
  DOI = {10.1038/s41467-023-43479-6},
  number = {1},
  pages={211},
  journal = {Nature Communications},
  publisher = {Springer Science and Business Media LLC},
  author = {Clinton,  Laura and Cubitt,  Toby and Flynn,  Brian and Gambetta,  Filippo Maria and Klassen,  Joel and Montanaro,  Ashley and Piddock,  Stephen and Santos,  Raul A. and Sheridan,  Evan},
  year = {2024},
}

@article{xianlong2006bethe,
  title = {Bethe ansatz density-functional theory of ultracold repulsive fermions in one-dimensional optical lattices},
  author = {Xianlong, Gao and Polini, Marco and Tosi, M. P. and Campo, Vivaldo L. and Capelle, Klaus and Rigol, Marcos},
  journal = {Phys. Rev. B},
  volume = {73},
  issue = {16},
  pages = {165120},
  numpages = {13},
  year = {2006},
  publisher = {American Physical Society},
  doi = {10.1103/PhysRevB.73.165120},
  url = {https://link.aps.org/doi/10.1103/PhysRevB.73.165120}
}

@article{yao,
  doi = {10.22331/q-2020-10-11-341},
  url = {https://doi.org/10.22331/q-2020-10-11-341},
  title = {Yao.jl: {E}xtensible, {E}fficient {F}ramework for {Q}uantum {A}lgorithm {D}esign},
  author = {Luo, Xiu-Zhe and Liu, Jin-Guo and Zhang, Pan and Wang, Lei},
  journal = {{Quantum}},
  issn = {2521-327X},
  publisher = {{Verein zur F{\"{o}}rderung des Open Access Publizierens in den Quantenwissenschaften}},
  volume = {4},
  pages = {341},
  month = oct,
  year = {2020}
}

@article{Zhang2018,
  title = {Quantum Algorithms to Simulate Many-Body Physics of Correlated Fermions},
  author = {Jiang, Zhang and Sung, Kevin J. and Kechedzhi, Kostyantyn and Smelyanskiy, Vadim N. and Boixo, Sergio},
  journal = {Phys. Rev. Appl.},
  volume = {9},
  issue = {4},
  pages = {044036},
  numpages = {23},
  year = {2018},
  publisher = {American Physical Society},
  doi = {10.1103/PhysRevApplied.9.044036},
  url = {https://link.aps.org/doi/10.1103/PhysRevApplied.9.044036}
}

@misc{NLopt,
  title = {The {NLopt} nonlinear-optimization package},
  author = {Steven G. Johnson},
  year = {2007},
  howpublished = {\url{https://github.com/stevengj/nlopt}}
}

@article{LBFGS,
  author = {Dong C. Liu and Jorge Nocedal},
  title = {On the limited memory {BFGS} method for large scale optimization},
  doi = {10.1007/bf01589116},
  year = {1989},
  volume = {45},
  pages = {503--528},
  journal = {Mathematical Programming},
  url = {http://dx.doi.org/10.1007/BF01589116},
}

@article{Cerezo2021,
author={Cerezo, M.
and Arrasmith, Andrew
and Babbush, Ryan
and Benjamin, Simon C.
and Endo, Suguru
and Fujii, Keisuke
and McClean, Jarrod R.
and Mitarai, Kosuke
and Yuan, Xiao
and Cincio, Lukasz
and Coles, Patrick J.},
title={Variational quantum algorithms},
journal={Nature Reviews Physics},
year={2021},
day={01},
volume={3},
number={9},
pages={625-644},
issn={2522-5820},
doi={10.1038/s42254-021-00348-9},
url={https://doi.org/10.1038/s42254-021-00348-9}
}

@article{Bharti2022,
  title = {Noisy intermediate-scale quantum algorithms},
  author = {Bharti, Kishor and Cervera-Lierta, Alba and Kyaw, Thi Ha and Haug, Tobias and Alperin-Lea, Sumner and Anand, Abhinav and Degroote, Matthias and Heimonen, Hermanni and Kottmann, Jakob S. and Menke, Tim and Mok, Wai-Keong and Sim, Sukin and Kwek, Leong-Chuan and Aspuru-Guzik, Al\'an},
  journal = {Rev. Mod. Phys.},
  volume = {94},
  issue = {1},
  pages = {015004},
  numpages = {69},
  year = {2022},
  publisher = {American Physical Society},
  doi = {10.1103/RevModPhys.94.015004},
  url = {https://link.aps.org/doi/10.1103/RevModPhys.94.015004}
}

@article{Tilly2022,
title = {The Variational Quantum Eigensolver: A review of methods and best practices},
journal = {Physics Reports},
volume = {986},
pages = {1-128},
year = {2022},
issn = {0370-1573},
doi = {10.1016/j.physrep.2022.08.003},
url = {https://www.sciencedirect.com/science/article/pii/S0370157322003118},
author = {Jules Tilly and Hongxiang Chen and Shuxiang Cao and Dario Picozzi and Kanav Setia and Ying Li and Edward Grant and Leonard Wossnig and Ivan Rungger and George H. Booth and Jonathan Tennyson}
}

@article{DDI,
author = {Hodgson, M. J. P. and Kraisler, Eli and Schild, Axel and Gross, E. K. U.},
title = {How Interatomic Steps in the Exact Kohn–Sham Potential Relate to Derivative Discontinuities of the Energy},
journal = {The Journal of Physical Chemistry Letters},
volume = {8},
number = {24},
pages = {5974-5980},
year = {2017},
doi = {10.1021/acs.jpclett.7b02615},
URL = {https://doi.org/10.1021/acs.jpclett.7b02615 },
}

@article{PhysRevResearch.5.013200,
  title = {Quantum computation for periodic solids in second quantization},
  author = {Ivanov, Aleksei V. and S\"underhauf, Christoph and Holzmann, Nicole and Ellaby, Tom and Kerber, Rachel N. and Jones, Glenn and Camps, Joan},
  journal = {Phys. Rev. Res.},
  volume = {5},
  issue = {1},
  pages = {013200},
  numpages = {22},
  year = {2023},
  publisher = {American Physical Society},
  doi = {10.1103/PhysRevResearch.5.013200},
  url = {https://link.aps.org/doi/10.1103/PhysRevResearch.5.013200}
}

@article{GGA,
  title = {Generalized Gradient Approximation Made Simple},
  author = {Perdew, John P. and Burke, Kieron and Ernzerhof, Matthias},
  journal = {Phys. Rev. Lett.},
  volume = {77},
  issue = {18},
  pages = {3865--3868},
  numpages = {0},
  year = {1996},
  publisher = {American Physical Society},
  doi = {10.1103/PhysRevLett.77.3865},
  url = {https://link.aps.org/doi/10.1103/PhysRevLett.77.3865}
}

@article{universal_review,
author = {Peverati, Roberto  and Truhlar, Donald G. },
title = {Quest for a universal density functional: the accuracy of density functionals across a broad spectrum of databases in chemistry and physics},
journal = {Philosophical Transactions of the Royal Society A: Mathematical, Physical and Engineering Sciences},
volume = {372},
number = {2011},
pages = {20120476},
year = {2014},
doi = {10.1098/rsta.2012.0476},

URL = {https://royalsocietypublishing.org/doi/abs/10.1098/rsta.2012.0476}
}

@Article{Marzari2021,
author={Marzari, Nicola
and Ferretti, Andrea
and Wolverton, Chris},
title={Electronic-structure methods for materials design},
journal={Nature Materials},
year={2021},
day={01},
volume={20},
number={6},
pages={736-749},
issn={1476-4660},
doi={10.1038/s41563-021-01013-3},
url={https://doi.org/10.1038/s41563-021-01013-3}
}

@article{lattice_under,
    author = {Tran, Fabien and Stelzl, Julia and Blaha, Peter},
    title = "{Rungs 1 to 4 of DFT Jacob’s ladder: Extensive test on the lattice constant, bulk modulus, and cohesive energy of solids}",
    journal = {The Journal of Chemical Physics},
    volume = {144},
    number = {20},
    pages = {204120},
    year = {2016},
    month = {05},
    issn = {0021-9606},
    doi = {10.1063/1.4948636},
    url = {https://doi.org/10.1063/1.4948636},
}

\end{document}
